\documentclass[fleqn,usenatbib]{mnras}

\usepackage[T1]{fontenc}

\DeclareRobustCommand{\VAN}[3]{#2}
\let\VANthebibliography\thebibliography
\def\thebibliography{\DeclareRobustCommand{\VAN}[3]{##3}\VANthebibliography}

\usepackage{graphicx}	
\usepackage{amsmath}	
\usepackage{amssymb}	
\usepackage{gensymb}
\usepackage{pdflscape}
\usepackage{siunitx}
\usepackage{newtxtext,newtxmath}

\title[Characterisation of eclipsing WDdM binaries]{Characterising eclipsing white dwarf M dwarf binaries from multi-band eclipse photometry}

\author[A.~J. Brown et~al.]{
Alex J. Brown,$^{1}$\thanks{E-mail: ajbrown2@sheffield.ac.uk (AJB)}
Steven G. Parsons,$^{1}$
Stuart P. Littlefair$^{1}$
James F. Wild,$^{1}$
R.~P.\ Ashley,$^{2}$
E. Breedt,$^{3}$
\newauthor
V.~S. Dhillon,$^{1,4}$
M.~J. Dyer,$^{1}$
M.~J. Green,$^{5}$
P. Kerry,$^{1}$
T.~R.\ Marsh,$^{6}$
I. Pelisoli,$^{6}$
D.~I. Sahman,$^{1}$
\\
$^{1}$Department of Physics and Astronomy, Hicks Building, The University of Sheffield, Sheffield, S3 7RH, UK\\
$^{2}$Isaac Newton Group of Telescopes, Apartado de Correos 321, Santa Cruz de La Palma, E-38700, Spain\\
$^{3}$Institute of Astronomy, University of Cambridge, Madingley Road, Cambridge CB3 0HA, UK\\
$^{4}$ Instituto de Astrofisica de Canarias, E38205 La Laguna, Tenerife, Spain\\
$^{5}$Department of Astrophysics, School of Physics and Astronomy, Tel Aviv University, Tel Aviv 6997801, Israel\\
$^{6}$Department of Physics, University of Warwick, Gibbet Hill Road, Coventry, CV4 7AL, UK
}

\date{Accepted XXX. Received YYY; in original form ZZZ}

\pubyear{2022}

\begin{document}
\label{firstpage}
\pagerange{\pageref{firstpage}--\pageref{lastpage}}
\maketitle

\begin{abstract}
With the prevalence of wide-field, time-domain photometric sky surveys, the number of eclipsing white dwarf systems being discovered is increasing dramatically. An efficient method to follow these up will be key to determining any population trends and finding any particularly interesting examples. We demonstrate that multi-band eclipse photometry of binaries containing a white dwarf and an M~dwarf can be used to determine the masses and temperatures of the white dwarfs to better than 5 per cent. For the M~dwarfs we measure their parameters to a precision of better than 6 per cent with the uncertainty dominated by the intrinsic scatter of the M~dwarf mass-radius relationship. This precision is better than what can typically be achieved with low-resolution spectroscopy. The nature of this method means that it will be applicable to LSST data in the future, enabling direct characterisation without follow-up spectroscopy. Additionally, we characterise three new post-common-envelope binaries from their eclipse photometry, finding two systems containing hot helium-core white dwarfs with low-mass companions (one near the brown dwarf transition regime) and a possible detached cataclysmic variable at the lower edge of the period gap.

\end{abstract}




\section{Introduction}
The majority of stars within binaries will evolve as if they were single stars, never interacting with their companion other than gravitationally. Around 25 per cent, however, are born sufficiently close that at some point during their lives they will interact, transferring material between them and potentially affecting their future evolution \citep{Willems2004}. Many of these close binary systems will undergo a phase in their lifetimes known as common-envelope evolution, where both stars orbit within a shared envelope of material drawn from the expanding outer layers of the more evolved star. Drag forces between the common-envelope and the two stars cause them to spiral in to shorter orbital periods. Assuming the binary doesn't merge during this phase, the immediate product will be a short period post-common-envelope binary (PCEB), usually containing a low-mass main~sequence star -- otherwise known as an M~dwarf -- and the remnant core of the more evolved star which will become a white dwarf (WD).

As well as being a key tracer of the relatively poorly understood common-envelope phase, these white dwarf main~sequence (WDMS) binaries are thought to be the progenitors to a wide variety of interesting and exotic astrophysical phenomena, from cataclysmic variables (CVs) and hot subdwarf stars \citep{Han2002} to the future gravitational-wave source double WDs and the cosmologically-important Type~Ia supernovae.

Additionally, the compact nature of these WDMS binaries -- specifically those with lower mass companions and referred to hereafter as WD+M~dwarf (WDdM) binaries -- means a relatively large proportion are seen to eclipse, allowing for the determination of precise constraints on the physical parameters of the system \citep{Parsons2017a, Parsons2018}. They are therefore ideal systems with which to test models of stellar physics as well as providing much-needed insight into the common-envelope phase itself \citep{Zorotovic2010, Toonen2013}. At the last published count, the sample of eclipsing WDdM PCEBs stood at around 80 systems \citep{parsons2015}. With large-scale photometric sky surveys such as the Zwicky Transient Facility (ZTF) \citep{Bellm2019, Graham2019}, and Legacy Survey of Space and Time (LSST) \citep{Ivezic2019} in the future, this number is set to increase considerably (with more than 200 systems already found in ZTF, \citealt{vanRoestel2019}). With fainter systems being discovered, this will make efficient and reliable follow-up and characterisation of these systems more difficult. Previously, follow-up observations of newly discovered WDdM systems have typically relied on low-resolution spectroscopy to determine initial system parameters, attempting to fit the Balmer sequence of the WD as well as the spectral energy distributions (SEDs) of both components \citep{Rebassa-Mansergas2007, Rebassa-Mansergas2010, Rebassa-Mansergas2013, Rebassa-Mansergas2016}. This is complicated by the dilution of spectral features due to the companion star which makes disentangling the two component spectra difficult and is hard to apply to the increasing population of fainter systems. Purely photometric approaches have been used recently e.g. \citet{Rebassa-Mansergas2021} who derived parameters for WDdM binaries using the Virtual Observatory SED Analyser to fit two component models to the SEDs and parallaxes of the systems. While this is a useful method, especially for large samples, it remains relatively untested and the uncertainties are difficult to estimate. It is also unclear how the fitted parameters are affected by any phase dependence on the photometry such as reflection effect or ellipsoidal modulation, especially given that the photometric measurements taken across the different bands are unlikely to be taken at similar orbital phases.

For eclipsing systems in particular, the shape of the eclipse light curve provides powerful constraints on the system parameters. However, this has not yet been exploited for initial parameter estimation, instead being used in conjunction with radial velocity measurements to retrieve precise model-independent parameters for detailed studies \citep{Parsons2010, Parsons2017a, Parsons2018}.
Given that a primary eclipse (the eclipse of the WD by the M~dwarf companion) light curve will require a similar amount of telescope time as a low-resolution identification (ID) spectrum, there is potential to use eclipse photometry for initial follow-up instead of low-resolution spectroscopy.
There are many benefits to this, one of which being that photometry can be used with fainter systems than spectroscopy can manage. This will become a major advantage as many of the systems discovered by LSST will be so faint that eclipse photometry will be the only viable route to measuring their parameters. Other benefits originate from the clean separation of the two stars, permitted by the eclipse of the WD. This removes the issue of disentangling the two component spectra that spectroscopy suffers from and allows for robust fitting of both stars even when one is much brighter than the other, especially useful for discerning systems with faint brown dwarf companions or small, high-mass WDs. Additional advantages come from the high temporal resolution of an eclipse light curve, allowing for detection of short-term variability which may indicate the presence of a WD that is pulsating or magnetic. These specific WDdM subtypes are of particular interest and would likely be missed by spectroscopy. Moreover, should an interesting system be discovered, eclipse light curve data can then be reused in combination with radial velocity measurements for more detailed study whereas low-resolution spectroscopy is much less useful beyond the initial parameter estimation.

Here, we present a Markov Chain Monte Carlo (MCMC) code\footnote{\url{https://github.com/Alex-J-Brown/pylcurve}} developed to make use of these advantages to fit the parameters of these systems, namely the masses and temperatures of the WD and M~dwarf components (referred to as the primary and secondary respectively), using only high-cadence multi-band photometry of the primary eclipse in combination with {\it Gaia} parallax measurements \citep{GaiaCollaboration2016, GaiaCollaboration2021} and theoretical models. We aim to do this more accurately and reliably than low-resolution spectroscopy and with sufficient precision to discern systems of particular interest. Specifically, we aim to determine the WD parameters to better than 5 per cent which is sufficient to discern interesting WD subtypes such as ZZ Cetis. For the M~dwarfs, we aim for a similar level of precision which is adequate for distinguishing companions at, or close to, the substellar boundary.
In terms of accuracy, the goal is for the parameters of both components to be within, at most, three standard deviations of the `true' values. Here we assume the high-precision model-independent values \citep{Parsons2017a, Parsons2018} to represent these `true' parameters.

\section{Eclipse modelling}

We construct a multi-wavelength light curve model using the \texttt{lcurve} code \citep[see][appendix A]{Copperwheat2010} to model the light curves of WDdM binaries immediately around the primary eclipse. The eclipse provides the strongest constraints on the system parameters and is sufficient to characterise the system without needing to observe a full orbit. Using this small region around the eclipse has the benefit of being much more efficient in terms of telescope time as well as being much faster to fit (due to a significant reduction in the number of light curve points that have to be modelled).

An \texttt{lcurve} eclipse model for a detached binary is defined by 19 parameters:
\begin{enumerate}
    \item The mass ratio, $q=\frac{M_{2}}{M_{1}}$.
    \item The radii, $R$, of each star scaled by the orbital separation, $a$, $\frac{R}{a}$. These radii are measured from the centre of mass of the star towards the inner Lagrangian point, $L_{1}$.
    \item The orbital inclination, $i$.
    \item The equivalent blackbody temperature, $T_{\mathrm{BB}}$, of each star. These blackbody temperatures, together with the effective wavelength, $\lambda_{\mathrm{eff}}$, of the model define the monochromatic flux normal to the surface of the star via the Planck law.
    \item The orbital ephemeris of the system, defined by the orbital period, $P$, and the time of mid-eclipse, $T_{0}$.
    \item The limb-darkening coefficients of both stars, $c_{1}$, $c_{2}$, $c_{3}$, and $c_{4}$, using the four-parameter prescription \citep{Claret2000}. 
    \item The bandpass-specific gravity-darkening coefficient for the secondary star, $y$.
    \item The fraction of incident flux from the WD that is absorbed by the M~dwarf, $F_{\rm{abs}}$.
\end{enumerate}

Many of these parameters vary with wavelength, most of which have little effect on the eclipse profile, resulting in an impractical number of free parameters when fitting multiple bands simultaneously. Additionally, degeneracies exist between some of these parameters, most notably the two scaled radii and the orbital inclination \citep{Parsons2017a}. Theoretical models and relations are therefore required in order to define these parameters from those that we are interested in -- the WD masses and effective temperatures and the masses of their M~dwarf companions -- during the fitting procedure. In the following sections we outline how this is achieved.

\subsection{Mass-radius relations}

The shape of the WD eclipse primarily constrains the scaled radii of the two stars and their orbital inclination. In order to retrieve masses from the eclipse photometry and break the degeneracy between scaled radii and inclination, mass-radius relations for both stars are required.

For the WD we use the mass-radius relations of \citet{Panei2007}, \citet{Fontaine2001}, and \citet{Althaus2005} for He-, CO-, and ONe-core compositions respectively; This core composition must be selected for a particular fit.\footnote{For systems with best-fit WD masses close to a border between core compositions it is worth running the fit again with the alternative core composition in order to determine which is most consistent with the expected mass range for the respective core composition (i.e. a $0.3~\rm{M}_{\odot}$ CO-core WD is unlikely).} Mass-radius relations for low- and intermediate-mass WDs have been well tested observationally and shown to be robust and accurate to within a couple of per cent \citep{Parsons2017a}. For higher mass WDs, the models are assumed to be similarly reliable, however, should this not be the case they will at least be sufficient to mark the system as containing a high mass WD worthy of further study.
In terms of the outer hydrogen layer mass, models with thick hydrogen layers ($M_{\textrm{H}}/M_{\textrm{WD}}=10^{-4}$) have been shown to represent WDs in PCEBs well in the majority of measured cases \citep{Parsons2017a}. We therefore use these thick layer models for our WD mass-radius relations.

\begin{figure}
 \includegraphics[width=\columnwidth]{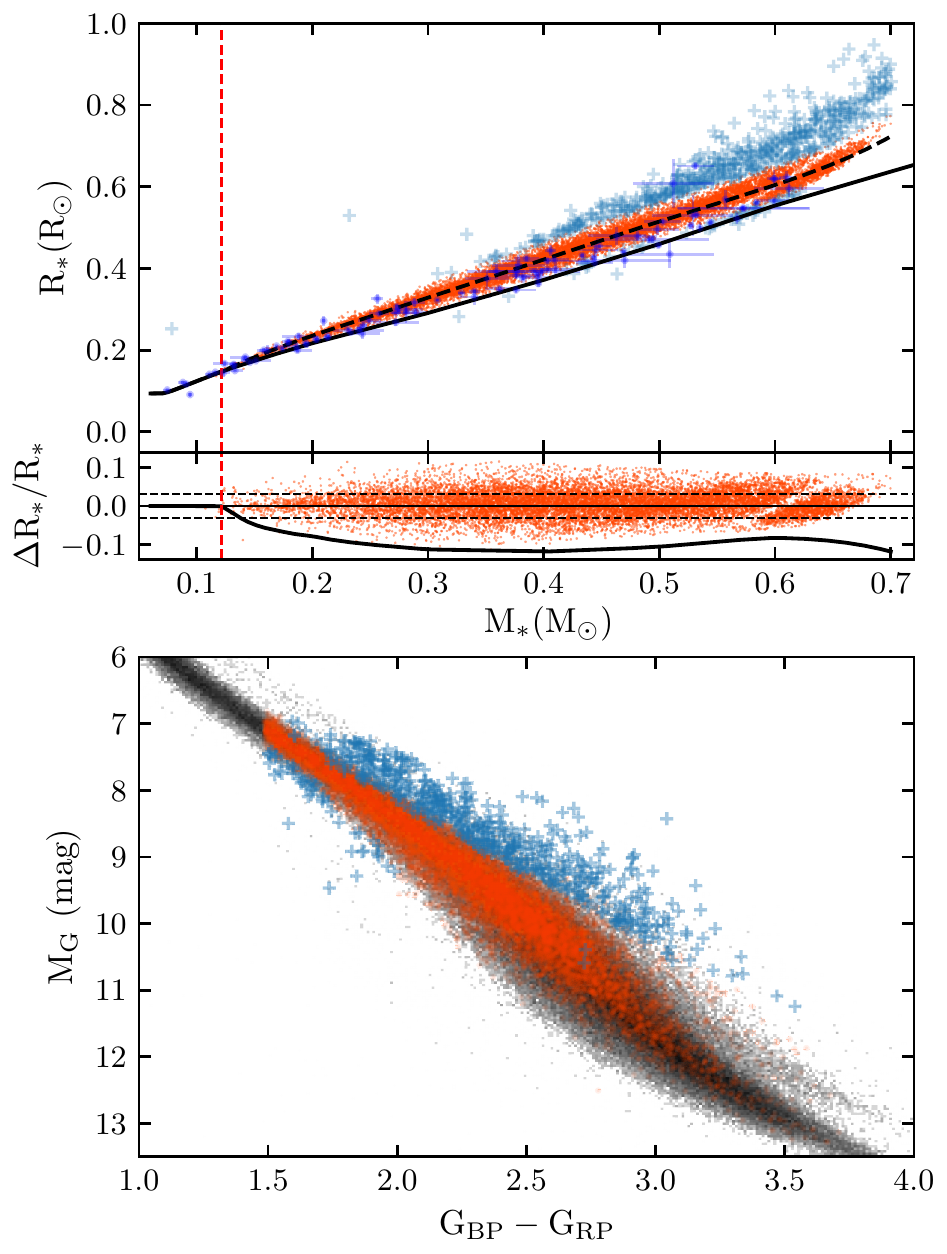}
 \caption{\textbf{above}) Semi-empirical M~dwarf mass-radius relation (black dashed line). Red points are those that remain after the sigma clipping while the light blue crosses are those that are discarded which are mainly pre-main-sequence stars or unresolved binaries. Dark blue points with error bars are M~dwarfs with well constrained masses and radii collated in \citet[table A1]{Parsons2018}. The solid black line shows the 1~Gyr track of \citet{Baraffe2015}. Fractional residuals relative to the semi-empirical relation are shown below with the dashed lines indicating $\pm 1\sigma$. The transition between the fitted relation and the Baraffe 1~Gyr model is indicated by the vertical dashed line. The gap in the data around $M_{*}=0.6~M_{\sun}$ is due to the discontinuity mentioned by \citet{Morrell2019} at $T_{\mathrm{eff}}=4000~K$.
 \textbf{below}) {\it Gaia} Hertzprung-Russell diagram for stars within 100~pc with the M~dwarf mass-radius sample overplotted in red. Blue crosses show those discarded by the sigma-clipping procedure demonstrating that they mostly lie above the main~sequence.}
 \label{fig:mdwarf_mr}
\end{figure}

M~dwarfs, however, are often found to be inflated relative to theoretical models for their mass, with radii typically found to be around 5 to 15 percent larger than models predict \citep{Lopez-Morales2005, Lopez-Morales2007, Parsons2018, Kesseli2018}. In an attempt to minimise any systematic effects arising from inflation, we produce a semi-empirical mass-radius relation for M~dwarfs.
We assign masses to a sample of 15\,279 M~dwarfs with {\it Gaia} parallaxes, radii, and 2MASS $K_{S}$ measurements \citep{Morrell2019} using the preferred fifth order ($n=5$) polynomial representation of the $K_{\rm{abs}}\textrm{--}M_{*}$ relation \citep{Mann2019}.
The resulting M~dwarf mass-radius relationship is shown in \autoref{fig:mdwarf_mr}. A population of stars exist above the main group. Checking these against the rest of the sample it is clear that they lie above the {\it Gaia} main~sequence and are likely binaries or pre-main-sequence stars, explaining their anomalous radii measurements. Cross-matching these outlying points with Simbad confirms that a large proportion of these are indeed pre-main-sequence stars, variables, or binaries and can be discarded.
We use an iterative sigma-clipping fitting procedure using a fifth-order polynomial to discard these points, removing $\approx3$ per cent of the total sample. We then follow this up with an MCMC fit to retrieve the final relation (\autoref{tab:mr_poly_coeffs}) while providing reliable uncertainty estimates on the polynomial coefficients. Due to the sparse nature of the sample in the low mass range and the convergence with the theoretical tracks of \citet{Baraffe2015}, we switch to using the theoretical models below the mass where the semi-empirical fit crosses the models. This occurs at $M_{*} = 0.121~\rm{M}_{\odot}$. We use the 1~Gyr model from \citet{Baraffe2015} below this point. We note that the apparent upturn in the sample above $\approx0.65~\rm{M}_{\odot}$ is not real and is a result of the fitted effective temperatures in the M~dwarf sample being limited to below $4400~\rm{K}$ and therefore stars with slightly higher temperatures than this will require a larger radius to fit their observed luminosity. We therefore only consider the fitted relation valid below this point.

There is some scatter in the M~dwarf sample around the best-fit semi-empirical relation. Much of this scatter is likely genuine variation in the radii of M~dwarfs with similar masses. This has been seen before with \citet{Parsons2018} measuring a scatter of $\approx5$ per cent in the radii of M~dwarfs in their sample. Additionally, some may be due to scatter in the fitted temperature for a given $K_{S}$ magnitude. This is demonstrated by the gap in the sample due to the discontinuity in models at $T_{\mathrm{eff}}=4000~K$, described by \citet{Morrell2019}, and which therefore describes a line of constant temperature. Additional contributions to the scatter come from $K_{S}$ magnitude uncertainties (typically below 2 per cent) and metallicity dependence (estimated to be $\approx1.7$ per cent, \citealt{Morrell2019}). 

The fractional residuals have a standard deviation of $\approx3$ per cent. This is slightly higher when measuring the scatter as a function of radius with fractional residuals of $\approx3.5$ per cent. This means that a fit with a hypothetical perfect determination of the secondary radius would translate into a secondary mass distribution with a standard deviation of 3.5 per cent and is therefore the maximum precision possible on the secondary mass using this relation. Additional errors in this mass-radius relation may be introduced through the $K_{\rm{abs}}\textrm{--}M_{*}$ relation used to derive it, with \citet{Mann2019} predicting that it is able to determine the mass of a star to a precision of $\approx2$ per cent. They also mention that there exists a small (\la2 per cent) systematic offset for literature M~dwarfs in eclipsing binary systems as compared to their predictions from the relation, possibly due to magnetic activity or rotation rates. Summing these contributions in quadrature with the 3.5 per cent scatter gives an estimated maximum precision in the secondary mass of $\approx5$ per cent. Any additional systematic contributions due to binarity (i.e. magnetic activity, rotation effects, or Roche distortion) are difficult to examine and we assume them to be small.

While this estimated uncertainty is straightforward to fold into the MCMC fitting routine, it increases the MCMC convergence time considerably, making it prohibitively long. We therefore choose to assume the best fit semi-empirical relation and account for the additional uncertainty in the relations at the end of the fitting, combining the formal errors from the MCMC in quadrature with the 5 per cent uncertainty for the secondary mass, and a 1 per cent uncertainty for the primary mass (as WD mass-radius relations have not yet been tested to higher precision than this).

\begin{table}
    \centering
    \begin{tabular}{lSSSSSS}
    \hline
         & $a_{5}$ & $a_{4}$ & $a_{3}$ & $a_{2}$ & $a_{1}$ & $a_{0}$ \\
        \hline
        $\mu$ & 27.4 & -54.1 & 41.1 & -14.9 & 3.53 & -0.124 \\
        $\sigma$ & 1.8 & 3.6 & 2.7 & 0.9 & 0.16 & 0.010 \\
        \hline
    \end{tabular}
    \caption{Best-fit coefficients and uncertainties for a fifth-order polynomial fit to the semi-empirical M~dwarf mass and radius measurements of the form $\frac{R}{R_{\odot}}=\sum_{n=0}^{5}a_{n}\left(\frac{M}{M_{\odot}}\right)^{n}$.}
    \label{tab:mr_poly_coeffs}
\end{table}

\subsection{Irradiation}
Many of these WDdM binaries contain hot WDs ($T_{\textrm{eff}}>20\ 000~\rm{K}$). Given the small orbital separations in PCEBs this can result in high irradiating fluxes incident on the surface of the M~dwarf, often many times greater than the typical outgoing flux from the secondary. This high irradiation can induce an inflation in the M~dwarf companion by effectively blocking outgoing flux over a portion of the star's surface area and therefore requiring a larger unirradiated surface (at most $\approx7$ per cent larger, \citealt{Ritter2000}) in order to expel the excess luminosity and retain thermal equilibrium. We attempt to include this effect in our model using a simplified method where we assume that the effective temperature over the full surface of the secondary is uniform. We calculate the effective surface area, $s_{\rm{eff}}$, over which the outgoing flux is blocked \citep[equation 60]{Ritter2000} which can then be used to determine the inflated radius, $R_{\rm{irr}}$, using
\begin{equation}
    R_{\rm{irr}} = R_{0}(1-s_{\rm{eff}})^{-0.1}
\end{equation}
where $R_{0}$ is the radius of the secondary without irradiation (i.e. the output from the semi-empirical mass-radius relation).

\subsection{Roche distortion}
The radii referred to in the previous sections are of an isolated and therefore spherically symmetric star. The compact nature of PCEBs mean that the Roche distortion of the secondary due to the WD can become significant and therefore needs to be corrected for. To do this we assume that the radius of a spherically symmetric star, used above, is equivalent to the volume averaged radius of the star when experiencing Roche distortion.
There is no analytical equation for calculating this correction. We therefore produce tables relating the scaled radius measured towards $L_{1}$, as used by \texttt{lcurve}, to the scaled radius of a spherically symmetric star with the equivalent volume as a function of binary mass ratio. For a given mass ratio the scaled radius towards $L_{1}$ defines a Roche equipotential representing the surface of the star. We then determine the positions of points on this equipotential surface over a range of latitudes and longitudes and compute the volume of the convex hull defined by these points. The volume-averaged scaled radius can then be easily determined. The conversion then becomes a simple interpolation given the binary mass ratio and the volume-averaged scaled radius.

\subsection{Blackbody temperatures}
As previously mentioned, the temperatures used by \texttt{lcurve} are a substitute for the monochromatic specific intensity normal to the surface of the star, i.e. at $\mu=\cos{\theta}=1$ where $\theta$ is the angle between the line normal to the stellar surface and the line of sight. \citet{Claret2020a} provide tables of specific intensities at $\mu=1$, together with limb-darkening coefficients, $c_{k}$, for WDs in both the SDSS \citep{Fukugita1996} and Super-SDSS \citep{Dhillon2021} photometric systems.

For main~sequence stars, no such tables exist for the Super-SDSS system. We therefore use the PHOENIX specific intensity model spectra from the G{\"o}ttingen Spectral Library \citep{Husser2013} to compute these. We calculate synthetic fluxes, $\langle f_{\lambda}\rangle_{x}$, in the Super-SDSS system according to 

\begin{equation}
    \label{eqn:syn_phot}
    \langle f_{\lambda}\rangle_{x} = \frac{\int f_{\lambda}(\lambda)S_{x}(\lambda)\lambda d \lambda}{\int S_{x}(\lambda)\lambda d \lambda},
\end{equation}
where $f_{\lambda}(\lambda)$ is the spectral flux density as a function of wavelength, $\lambda$, and $S_{x}(\lambda)$ is the throughput of the chosen filter.
We do this at each value of $\mu$ supplied by the PHOENIX spectra. These can then be normalised to the flux at $\mu=1$ and fit with the four-parameter law of \citet{Claret2000},

\begin{equation}
    \label{eqn:limbdarkening}
    \frac{I_{\lambda}(\mu)}{I_{\lambda}(1)} = 1-\sum_{k=1}^{4}c_{k}(1-\mu^{\frac{k}{2}}),
\end{equation}
where $I_{\lambda}(\mu)$ is the specific intensity relative to that at $\mu=1$.
Rather than use the synthetic fluxes at $\mu=1$ to determine the \texttt{lcurve} temperatures, we calculate the total synthetic flux of the star, $F_{\lambda}$, according to our best-fit limb-darkening law using 

\begin{equation}
    \label{eqn:int_disk}
    F_{\lambda} = 2 \pi \int^{1}_{0} I_{\lambda}(\mu)\mu d\mu.
\end{equation}
We then take the central intensity required to match the total synthetic flux with the synthetic flux calculated for the equivalent HiRes PHOENIX spectrum. This ensures the absolute flux of a star modelled with these limb darkening parameters remains consistent with the full disk PHOENIX model.

Temperatures of blackbody spectra that give monochromatic specific intensities equal to these specific intensities at $\mu=1$ are then computed at the pivot wavelength of each filter in the SDSS and Super-SDSS systems.

\subsection{Model summary}

In summary, multi-band eclipse light curves clearly resolve the SEDs\footnote{When referring to the method described in this work, the SED is from the eclipse light curves alone and does not include any additional photometric data} of both components, with the depths of the eclipses showing the flux contributed by the WD and the in-eclipse flux showing the contribution from the secondary. These SEDs from the eclipse light curves constrain the effective temperatures of both stars which, together with parallax information, places constraints on their radii. The shape of the eclipse strengthens this constraint whilst also restricting the orbital separation and therefore the masses of the stars when combined with mass-radius relations.

We use \texttt{lcurve} to model these light curves, defining our \texttt{lcurve} model from the parameters of interest -- $T_{1}$, $T_{2}$, $M_{1}$, $M_{2}$, $i$, $T_{0}$, $\varpi$, and $E(B-V)$, where $\varpi$ is the parallax -- together with the orbital period, $P$, and the bandpass of the observation, via various theoretical models and relations. The \texttt{lcurve} parameters, described previously, are defined as follows:

\begin{enumerate}
    \item The mass ratio, $q$, is set from the masses as $\frac{M_{2}}{M_{1}}$.
    \item The scaled radius of the primary, $\frac{R_{1}}{a}$, is defined by the WD mass-radius relation for the chosen core composition together with Kepler's third law and therefore depends on the WD mass and temperature, the secondary mass, and the orbital period.
    The scaled radius of the secondary, $\frac{R_{2}}{a}$, is primarily defined by the semi-empirical mass-radius relation together with Kepler's third law, with corrections for irradiation and Roche distortion. It is therefore dependent on the WD mass and temperature, the secondary mass and temperature, and the orbital period of the system.
    \item The orbital inclination, $i$, is a free parameter.
    \item The equivalent blackbody temperatures of each star, $T_{\mathrm{BB}}$, is defined by the mass and temperature of the respective star together with the chosen bandpass.
    \item For the orbital ephemeris, $T_{0}$ is a free parameter in the fit while $P$ is fixed at a previously determined value.
    \item The limb-darkening coefficients of both stars, $c_{1}$, $c_{2}$, $c_{3}$, and $c_{4}$, like the blackbody temperatures, are defined by the mass and temperature of the respective star together with the chosen bandpass.
    \item The gravity-darkening coefficient for the secondary star, $y$, is also defined by its mass and temperature along with the chosen bandpass.
    \item The fraction of incident flux from the primary that is absorbed by the secondary can generally be ignored due to only fitting a small region surrounding the primary eclipse. We therefore leave it fixed at $F_{\rm{abs}}=0.5$.
\end{enumerate}

\subsection{$\chi^{2}$ calculation for flux calibrated light curves}
\label{sec:chisqr}

When generating a model light curve, \texttt{lcurve} can be supplied with a scale factor which sets the absolute flux level of the light curve. It is therefore possible to calculate the scale factor required to produce a true flux light curve model from the parallax, interstellar extinction, and orbital separation. The issue with this approach is that any small error in the flux calibration of the data will cause issues with the fitting. This is because the flux calibrated eclipse light curves are unlikely to correspond exactly to the model SEDs of both stars, resulting in the fit being unable to correctly model both the in-eclipse and out-of-eclipse flux simultaneously, preventing an accurate fit to the eclipse shape and therefore reliable parameter estimation. We instead allow \texttt{lcurve} to automatically scale the model to the data, calculating the $\chi^{2}$ using this scaled model.

To include the absolute flux information, we take the WD flux contribution output by \texttt{lcurve} for the scaled model (a reliable measure of the depth of the primary eclipse) and compare this to the theoretical WD flux for the given temperature, mass, parallax, and extinction, calculating the $\chi^{2}$ for this using the flux calibration uncertainty. This method allows for a more proper handling of the uncertainties, treating those from the flux calibration and from the differential photometry independently. We combine these two values of $\chi^{2}$, repeating this over all observed bands to obtain an overall $\chi^{2}$ value for the full flux calibrated model.

\subsection{Fitting procedure}
To fit the light curves of a system we use MCMC, implemented through the \texttt{emcee} Python package \citep{Foreman-Mackey2013}. For each walker position an \texttt{lcurve} light curve model is generated for each observed bandpass with the log probability calculated as described in Section \ref{sec:chisqr}. The log probability from this model is combined with a parallax prior. We use a bounded Gaussian prior with a mean and standard deviation corresponding to the {\it Gaia} \texttt{parallax} and \texttt{parallax\_error} \citep{GaiaCollaboration2021} respectively with a limit at two standard deviations above and below.

The use of the {\it Gaia} parallax distribution as a prior in the MCMC prevents systematic issues that would be introduced by using the distance. All systems considered in this work have a $\texttt{parallax\_over\_error} > 10$ and, when combined with the photometric information, will be constrained sufficiently that the inclusion of the galactic stellar density distribution is unnecessary.

For systems with larger uncertainty in their parallax measurement, or maybe even no parallax information at all as will be the case for many of the systems discovered by LSST, this method can still be successful in fitting the parameters. It is possible, however, that in the case of systems where the SED of either star does not match the models well (most likely due to irradiation effects or star spots on the secondary), that the fit will compensate using the parallax. For systems with good parallax measurements, this effect is relatively obvious and can be flagged. Assuming it is the secondary SED that is causing the problem (which is most likely) then allowing its temperature to be independent in each band can allow the fit to converge to values consistent with the measured parallax\footnote{Note that this increases the number of dimensions for the MCMC and so increases the convergence time as well as removing the effective temperature information for the secondary. It is therefore best left as a backup method in the case where the original fit is struggling.}. For those with high parallax uncertainties, though, it may go unnoticed, leading to incorrect parameters. 

Priors on all other parameters are uniform with upper and lower limits defined by the range of the model grids.
Each MCMC chain is run with 100 walkers for 20\,000 steps. The chains are then inspected and a number of steps are discarded from the beginning as burn-in. The number of steps discarded is chosen to remove any steps before the final equilibrium position.

\section{Observations}
\label{sec:observations} 
In order to validate this modelling, we test our fitting code on three previously well characterised WDdM PCEBs. We use archival photometry of these systems from ULTRACAM \citep{Dhillon2007} and HiPERCAM \citep{Dhillon2021} to test that the method is successful when using either three-band or five-band data. We then fit three, previously unpublished, systems, all observed with ULTRACAM (all observations detailed in \autoref{tab:observations})
\begin{table*}
    \centering
    \begin{tabular}{cccccccc}
    \hline
        Target & Date & UT start & Filters & Telescope-Instrument & Exp time (s) & Sky transparency & FWHM (") \\
        \hline
        NN~Ser & 2019-07-09 & 23:39:06 & $u_{s}g_{s}r_{s}i_{s}z_{s}$ & GTC-HiPERCAM & 1.0 & Some dust but stable & 1.5 \\
        SDSS~J0838$+$1914 & 2010-12-13 & 07:41:55 & $u'g'r'$ & NTT-ULTRACAM & 4.8 & Photometric & 2 \\
        SDSS~J1028$+$0931 & 2018-01-30 & 06:07:53 & $u_{s}g_{s}r_{s}$ & NTT-ULTRACAM & 2.5 & Photometric & 1 \\
        2MASS~J1358$-$3556 & 2018-06-01 & 03:04:59 & $u_{s}g_{s}i_{s}$ & NTT-ULTRACAM & 4.0 & Photometric & 1.5 \\
        EC~12250$-$3026 & 2018-05-31 & 23:03:24 & $u_{s}g_{s}i_{s}$ & NTT-ULTRACAM & 3.0 & Photometric & 1.5 \\
        SDSS~J1642$+$0135 & 2019-03-04 & 07:40:28 & $u_{s}g_{s}i_{s}$ & NTT-ULTRACAM & 5.0 & Photometric & 1.5 \\
        \hline
    \end{tabular}
    \caption{Journal of observations.}
    \label{tab:observations}
\end{table*}

\subsection{Reduction}
We used the HiPERCAM pipeline \citep{Dhillon2021} to debias, flat-field correct -- and defringe in the case of HiPERCAM $z_{s}$ data -- and then extract aperture photometry. We allowed the radius of the target aperture to vary in line with the measured full-width at half-maximum of a reference star in each frame to minimise the effects of seeing variation. The counts from the target were measured relative to a brighter comparison star to remove any transparency variations and atmospheric extinction effects.

\subsection{Flux calibration}
Using the SED of the WD to determine its temperature requires precise flux calibration. This requirement is emphasised by the significant temperature dependence of the WD mass-radius relations on which this method relies heavily. Flux calibration is complicated by the significant departure of the HiPERCAM and ULTRACAM Super-SDSS filter sets from the standard SDSS system. This departure is most notable in the $u_{s}$ band where it can be tenths of magnitudes. The typical solution would be to observe a selection of spectro-photometric standard stars with well known spectra -- observed or theoretical -- spanning the full wavelength range of the filter set in order to calibrate the observed light curves from synthetic photometry. This is the method that will be employed in future works using the {\it Gaia} spectro-photometric standard stars \citep{Pancino2012, Altavilla2015, Marinoni2016, Altavilla2021, Pancino2021}. Synthetic AB magnitudes of these {\it Gaia} spectro-photometric standard stars computed for the ULTRACAM and HiPERCAM Super-SDSS systems are included in appendix \ref{section:appendix} for future reference along with a more thorough analysis of the differences between the Super-SDSS and SDSS photometric systems.

However, as this work is based on archival photometry, only flux standards with USNO--40 photometry \citep{Smith2002} are available for flux calibration. We instead fit PHOENIX model spectra to the USNO--40 and {\it Gaia} photometry of these standards to calculate the magnitude offsets required to transform the USNO--40 photometry into the Super-SDSS system.

Using an MCMC method, implemented through the \texttt{emcee} Python package, we fit the effective temperature, surface gravity, radius, interstellar reddening, and parallax of the standard stars. At each walker step the corrections between the two filter systems are saved. This gives us the ability to propagate the effects of any non-trivial correlations through to the uncertainties in the desired offsets. 
We run 100 walkers for 10,000 steps and discard the first 2000 as burn-in. 

We measure the atmospheric extinction in each filter by fitting a first order polynomial to the instrumental magnitudes as a function of airmass of any bright stars included in an observing run that covers a good airmass range on the same night as our target. We then use this, along with the transformed Super-SDSS standard star magnitudes, to calibrate the comparison star of our target. The target is therefore calibrated by performing differential photometry against this comparison star.

\subsection{Comparison with previously published systems}
\subsubsection{NN~Ser}
NN~Ser is a well-characterised eclipsing binary, first discovered by \citet{Haefner1989}, containing a hot WD and an M~dwarf companion with more than a decade of archival high-speed multi-colour photometry. \citet{Parsons2010} combined ULTRACAM photometry with phase-resolved UVES spectroscopy to obtain precise parameters for the system, independent of any mass-radius relations. This makes it an ideal system with which to test the purely photometric approach presented in this paper. Additionally NN~Ser has been observed with HiPERCAM allowing us to test the method with simultaneous $u_{s}g_{s}r_{s}i_{s}z_{s}$ data.

Initial fits to the HiPERCAM photometry struggled, attempting to change the parallax to values inconsistent with the {\it Gaia} measurement. We determined this was due to the M~dwarf SED (the in-eclipse HiPERCAM photometry) not matching the PHOENIX model for the published parameters, instead preferring a higher temperature model. The fit was compensating for this by altering the radius, and therefore mass of the secondary, having an effect on the rest of the system parameters. This likely reflects the fact that the high irradiation experienced by the M~dwarf prevents it from being well represented by the PHOENIX models, even on the unirradiated face seen during the primary eclipse. Allowing the temperature of the secondary to be independent in each band solved this issue, allowing the MCMC to converge to a fit consistent with the {\it Gaia} parallax.

Our best fit to the HiPERCAM photometry (see \autoref{tab:old_systems} and \autoref{fig:published_systems}) achieves uncertainties (and therefore precisions) in the WD mass, effective temperature, and secondary mass of 1.8, 4.6, and 5.7 per cent respectively. Although our secondary mass is $\approx11$ per cent more massive than the published value it is still consistent to better than $2\sigma$. The other parameters all lie within $1\sigma$ of the high precision, published values \citep{Parsons2017a, Parsons2018}. This demonstrates that, even for systems with highly irradiated companions, our method can still be successful.

\subsubsection{SDSS~J0838$+$1914}
SDSS~J0838$+$1914 (SDSS~J083845.86$+$191416.5 in SIMBAD, also known as CSS~40190) is another well characterised PCEB and was discovered by \citet{Drake2010} in the Catalina Sky Survey. It was later characterised by \citet{Parsons2017a, Parsons2018} who found it to contain a WD with a temperature and mass of $14\,900\pm730~\rm{K}$ and $0.482\pm0.008~\rm{M}_{\odot}$, respectively, in a $3.123~\rm{h}$ orbit with a $3100\pm100~\rm{K}$ main~sequence companion with a mass of $0.142\pm0.013~\rm{M}_{\odot}$. It is therefore a fairly typical example of a PCEB. Good quality archival ULTRACAM photometry is available for this system. It has also been characterised from SDSS spectroscopy making it a good system with which to compare our photometric fit against the spectroscopic one.

Fitting the ULTRACAM eclipse photometry of the system (see \autoref{tab:old_systems} and \autoref{fig:published_systems}) we achieve a precision in the WD mass and effective temperature of 2.9 and 2.4 per cent respectively and the secondary mass and effective temperature of 6.1 and 0.3 per cent respectively. Comparing these to the SDSS spectroscopic values which determine the WD mass and effective temperature to a precision of 9.0 and 3.0 per cent respectively and the secondary mass to 49 per cent demonstrates that we can reach a higher precision using eclipse photometry. Additionally all our best fit values lie within $2\sigma$ of the high-precision, published values \citep{Parsons2017a, Parsons2018}.

\subsubsection{SDSS~J1028$+$0931}
SDSS~J1028$+$0931 (SDSS~J102857.78$+$093129.8 in SIMBAD) was discovered to be an eclipsing WDdM system by \citet{parsons2013a}. Later and more precise characterisation was performed by \citet{Parsons2017a, Parsons2018}, who found a WD temperature and mass of $12\,221\pm765~\rm{K}$ and $0.4146\pm0.0036~\rm{M}_{\odot}$, respectively with the fit to the secondary giving a temperature and mass of $3\,500\pm100~\rm{K}$ and $0.403\pm0.005~\rm{M}_{\odot}$. The higher mass secondary in this system means that the spectrum is dominated by the M~dwarf redward of the $g$ band with measurable contribution in the $u$ band. This makes it a difficult system to characterise from low-resolution spectroscopy due to the dilution of the Balmer series of the WD by the M~dwarf. We fit this system in order to show that the presence of eclipses overcomes the issues caused by the superposition of both SEDs and results in a more robust fit.

\begin{figure*}
    \centering
    \includegraphics[width=\textwidth]{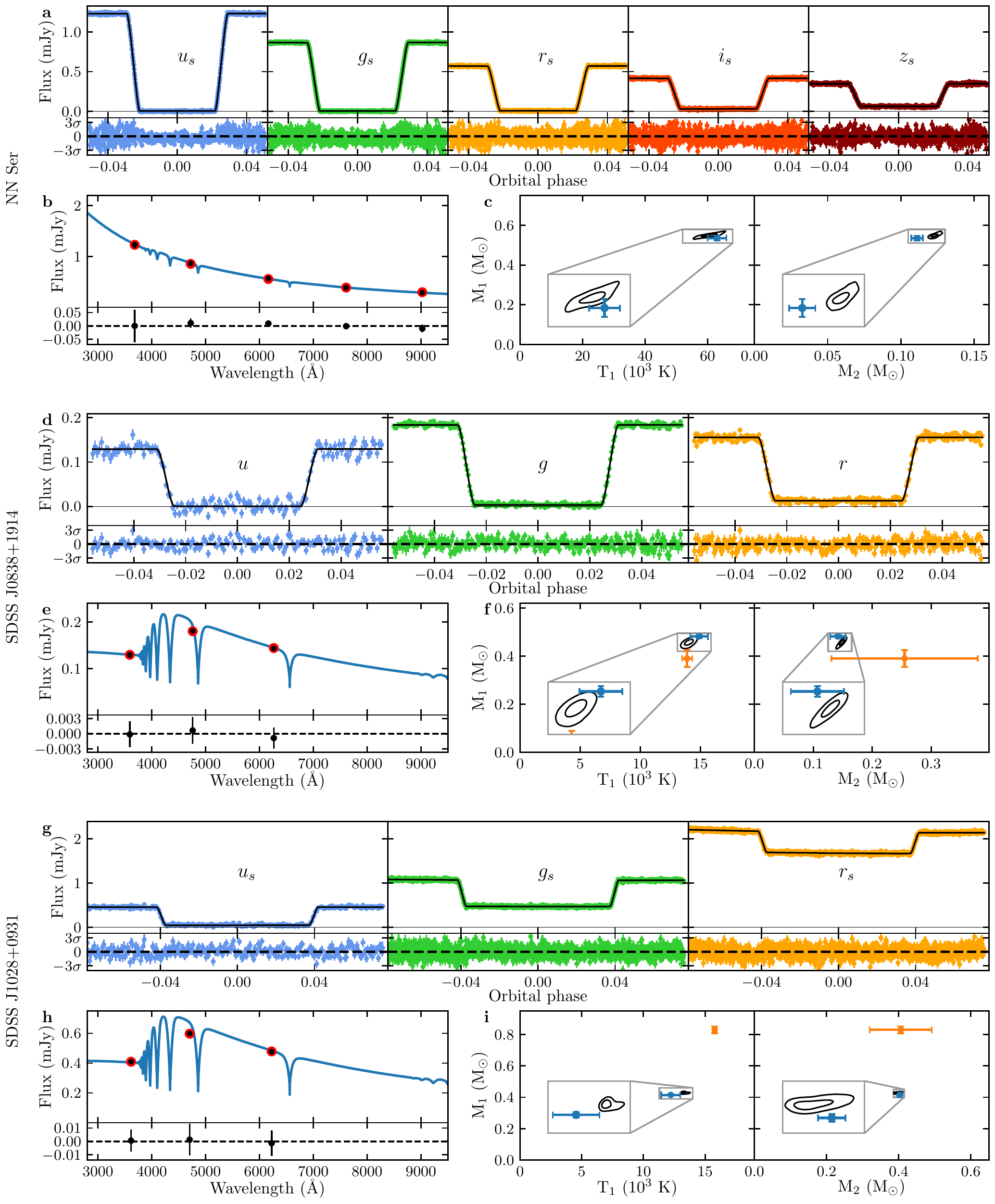}
    \caption{Best fit results for the three previously published systems. Light curve plots show the flux calibrated eclipse light curve data (coloured points) along with the best fit eclipse model (black line). A horizontal black line shows a flux of zero for reference with the residuals shown below. The panel below the light curves to the left shows the SED of the WD i.e. the depths of the eclipses in each band (black points). These are shown against the \citet{Koester2010} model spectrum for the best fit parameters (blue line) and the synthetic photometry from this model (red points) with the residuals below. The panel to the right shows the $1\sigma$ and $2\sigma$ contours from our MCMC fit along with the published parameters (blue points, \citealt{Parsons2017a, Parsons2018}) and those derived from SDSS spectroscopy where available (orange points, \citealt{Rebassa-Mansergas2012}).}
    \label{fig:published_systems}
\end{figure*}

The secondary of SDSS~J1028$+$0931 appears to have at least one significant star spot on its surface which adds an additional slope to the photometry. We add a linear term to the $g_{s}$ and $r_{s}$ band models to account for this. The best fit including these two additional parameters (see \autoref{tab:old_systems} and \autoref{fig:published_systems}) achieves an precision of 1.4 per cent on the WD mass and 1.9 per cent on its temperature. For the M~dwarf we manage 5.3 and 0.8 per cent respectively for the mass and temperature. Again, all of our best-fit parameters are within $2\sigma$ of the high-precision values of \citet{Parsons2017a, Parsons2018}. Comparing our fit with the spectroscopic values from SDSS demonstrates the benefit of eclipse photometry in systems where one component dominates, with the spectroscopic determination of the WD temperature being discrepant by over $4\sigma$ and the mass being discrepant by almost $10\sigma$.

\begin{table*}
    \centering
    \begin{tabular}{@{\extracolsep{4pt}}lcccccc@{}}
        \hline
         & \multicolumn{2}{c}{NN~Ser} & \multicolumn{2}{c}{SDSS~J0838$+$1914} & \multicolumn{2}{c}{SDSS~J1028$+$0931} \\
         \cline{2-3}
         \cline{4-5}
         \cline{6-7}
          & $\mu$ & Difference \%($\sigma$) & $\mu$ & Difference \%($\sigma$) & $\mu$ & Difference \%($\sigma$) \\
          \hline
        $T_{1}$(K) & $60800^{+2200}_{-2800}$ & $-3.5(-0.6)$ & $14060^{+340}_{-340}$ & $-5.7(-1.0)$ & $13270^{+250}_{-140}$ & $+8.6(+1.3)$ \\[0.9ex]
        $T_{2}$(K) & -- & -- & $2910^{+10}_{-10}$ & $-6.2(-1.9)$ & $3550^{+30}_{-20}$ & $+1.3(+0.4)$ \\[0.9ex]
        $M_{1}$($\rm{M}_{\odot}$) & $0.548^{+0.010}_{-0.009}$ & $+2.4(+0.9)$ &  $0.456^{+0.012}_{-0.013}$ & $-5.2(-1.7)$ & $0.428^{+0.006}_{-0.006}$ & $+3.2(+1.9)$ \\[0.9ex]
        $M_{2}$($\rm{M}_{\odot}$) & $0.123^{+0.007}_{-0.007}$ & $+10.8(+1.6)$ & $0.148^{+0.009}_{-0.009}$ & $+4.3(+0.4)$ & $0.397^{+0.021}_{-0.021}$ & $-1.4(-0.3)$ \\
        \hline
    \end{tabular}
    \caption{Comparison of parameters determined using our purely photometric method against published, model-independent values \citep{Parsons2017a, Parsons2018}. We show the deviation from the published values as a percentage and in units of standard deviation where the standard deviation is the uncertainty from our photometric fit summed in quadrature with the uncertainty of the published value. We also include a 1 per cent and 5 per cent systematic error contribution for the primary and secondary masses respectively.}
    \label{tab:old_systems}
\end{table*}
\subsection{New systems}

With the code proving successful for these three well-characterised systems, we now apply it to three previously uncharacterised PCEBs observed with ULTRACAM. Best-fit parameters are listed in \autoref{tab:new_systems} with the light curves shown in \autoref{fig:new_systems}.

\subsubsection{2MASS~J1358$-$3556}
2MASS~J1358$-$3556 (2MASS~J13581075$-$3556194 in the 2MASS catalogue or Gaia~DR2~6121651418527918976 in SIMBAD) was found to be an eclipsing PCEB by combining {\it Gaia} measurements and data from the Catalina Real-Time Transient Survey (CRTS) \citep{Drake2009}. Fitting the ULTRACAM eclipse photometry in $u_{s}$, $g_{s}$, and $i_{s}$ gives a WD mass of $0.438\pm0.007~\rm{M}_{\odot}$ with an effective temperature of $40\,600^{+1600}_{-2200}~\rm{K}$ assuming a helium core composition. Rerunning the fit with a carbon-oxygen core mass-radius relation favours a lower WD mass of $0.38~\rm{M}_{\odot}$. This mass is below what is expected for a WD with a CO-core composition and so the helium fit appears the most consistent. The M~dwarf in this system fits best with a mass of $0.118\pm0.006~\rm{M}_{\odot}$ and an effective temperature of $2980^{+30}_{-40}~\rm{K}$. The best fit parameters are shown in \autoref{tab:new_systems} with the light curve model shown in \autoref{fig:new_systems}. This system appears to be a fairly typical, if quite young, PCEB due to the high WD temperature.

\subsubsection{EC~12250$-$3026}
\begin{figure}
 \includegraphics[width=\columnwidth]{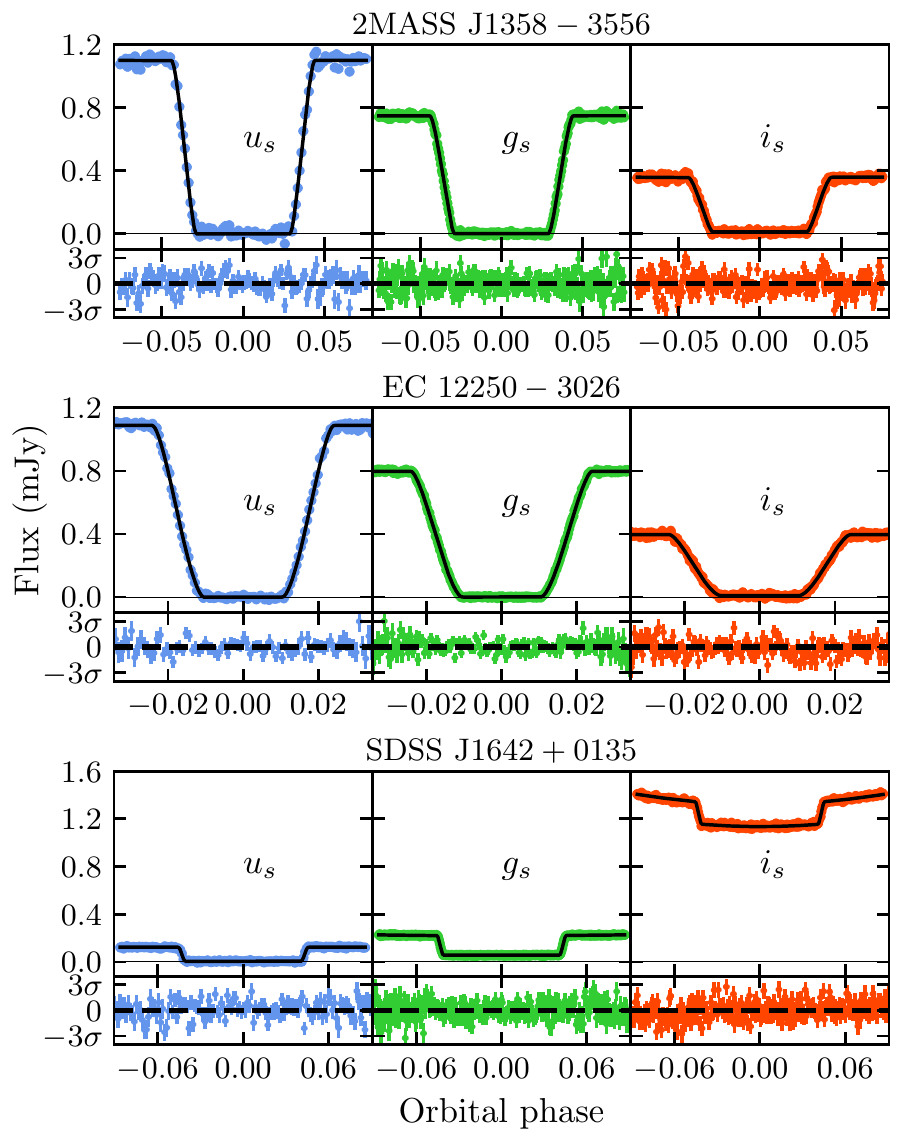}
 \caption{Best fits to ULTRACAM eclipse photometry of the three new systems. The best fit model is shown by the solid black line while a thin black line shows the zero level. Normalised residuals are shown below with a dashed black line showing zero.}
 \label{fig:new_systems}
\end{figure}

EC~12250$-$3026 was initially found in the Edinburgh-Cape blue object survey \citep{Stobie1997} where it was thought to be a single hot subdwarf. It was later found to be an eclipsing WDdM system in CRTS. The fit to the ULTRACAM photometry (\autoref{tab:new_systems}, \autoref{fig:new_systems}) gives a WD temperature of $33\,900^{+1000}_{-1300}~\rm{K}$ with a mass of $0.420^{+0.010}_{-0.009}~\rm{M}_{\odot}$, again favouring a helium core composition. The secondary in this system -- which is only detected in the $i_{s}$ band -- is quite low mass at $0.089\pm0.005~\rm{M}_{\odot}$ although still stellar. The low-mass of the secondary star makes this an interesting system for investigating the brown dwarf transition regime. For the secondary temperature, we find $2840^{+60}_{-30}~\rm{K}$ although it is worth mentioning that, like NN~Ser, the high temperature of this WD may mean that the best fit temperature of the M~dwarf is influenced by the high irradiation and is not necessarily representative of the true unirradiated temperature.

\subsubsection{SDSS~J1642$+$0135}

SDSS~J1642$+$0135 (SDSS~J164251.54$+$013554.9 in SIMBAD) was discovered by \citet{Denisenko2018} who determined it to be a pre-cataclysmic variable with a cool ($T_{\rm{eff}}\loa8000~\rm{K}$) white dwarf primary and strong ellipsoidal modulation. This system is of particular interest as its period of $2.31~\rm{h}$ places it in the cataclysmic variable (CV) period gap. Initial fits of this system were unable to match the extreme ellipsoidal modulation present in the $g_{s}$ and $i_{s}$ band photometry, even with a Roche lobe filling factor equal to one. This required adding the secondary gravity-darkening exponent, $\beta_{1}$ \citep{Claret2011}, to the model as a free parameter, allowing the fit to increase the effect of gravity darkening, preferring a value of $\beta_{1}=0.41\pm0.02$. This is roughly twice what would be expected for a star of the best fit temperature \citep{Claret2003}, however, observational validation of how $\beta_{1}$ varies with effective temperature seems to indicate a fairly large scatter around the theoretical values \citep[see figure 3]{Claret2003}.
With the addition of $\beta_{1}$ as a free parameter we confirm the predictions of \citet{Denisenko2018}, finding a $7650\pm60~\rm{K}$ WD with a mass of $0.69^{+0.010}_{-0.011}~\rm{M}_{\odot}$ and a secondary that is on the verge of -- if not already -- filling its Roche lobe with a linear filling factor of $0.97\pm0.017$. We find a secondary mass of $0.198\pm0.010~\rm{M}_{\odot}$ and effective temperature of $2897^{+5}_{-6}~\rm{K}$. The higher than average WD mass for PCEBs \citep{Zorotovic2011} is consistent with that of the volume-limited CV population \citep{Pala2020}. Additionally, the secondary mass matches the donor mass of $M_{\mathrm{donor}}=0.2\pm0.02~\rm{M}_{\odot}$ at which CV mass transfer ceases at the start of the period gap, as determined by \citet{Knigge2006}. It therefore seems possible that SDSS~J1642$+$0135 is a temporarily detached CV in the final stages of crossing the period gap with an orbital period $\approx10~\rm{mins}$ greater than the predicted period at the lower edge of the gap of $P=2.15~\rm{h}$ \citep{Knigge2006}. A population of apparently gap-crossing CVs has been identified statistically using the SDSS sample of PCEBs \citep{Zorotovic2016}, however, this may be one of the first specific examples of an eclipsing, gap-crossing CV and is worth more detailed study.

\begin{table*}
    \centering
    \begin{tabular}{ccccccccc}
    \hline
         Target & $T_{1}$~(K) & $T_{2}$~(K) & $M_{1}$~($\rm{M}_{\odot}$) &  $M_{2}$~($\rm{M}_{\odot}$) & $i$~($\degree$) & $E(B-V)$ & $T_{0}~(\rm{BMJD})$ & $P$~(d)\\
          \hline
        2MASS~J1358$-$3556 & $40600^{+1600}_{-2200}$ & $2980^{+30}_{-40}$ & $0.438^{+0.007}_{-0.007}$ & $0.118^{+0.006}_{-0.006}$ & $88.7^{+0.9}_{-1.3}$ & $0.04^{+0.01}_{-0.01}$ & $58270.1646820(12)$ & 0.0815296700(35) \\[0.9ex]
        EC~12250$-$3026 & $33900^{+1000}_{-1300}$ & $2840^{+60}_{-30}$ & $0.420^{+0.010}_{-0.009}$ & $0.089^{+0.005}_{-0.005}$ & $85.1^{+0.1}_{-0.2}$ & $0.04^{+0.01}_{-0.01}$ & $58269.97928329(67)$ & 0.1234482746(40) \\[0.9ex]
        SDSS~J1642$+$0135 & $7650^{+60}_{-60}$ & $2897^{+5}_{-6}$ & $0.693^{+0.010}_{-0.011}$ & $0.198^{+0.010}_{-0.010}$ & $89.1^{+0.6}_{-0.7}$ & $0.01^{+0.01}_{-0.01}$ & $58546.3505016(17)$ & 0.09629068740(52) \\
        \hline
    \end{tabular}
    \caption{Results of the MCMC fits to the three new WDdM systems.}
    \label{tab:new_systems}
\end{table*}

\section{Discussion}

Assessing the precision of our fits to the six systems, the median percentage uncertainty on the WD mass determination is 1.7 per cent with a maximum uncertainty of 2.9 per cent for SDSS~J0838+1914. For the  secondary mass these median and maximum uncertainties are 5.5 per cent and 6.1 per cent respectively which is dominated by the 5 per cent estimated contribution from systematic errors arising from the intrinsic scatter of the M~dwarf mass-radius relationship. This precision is, therefore, at or below the aim of 5 per cent precision which is necessary to discern systems with interesting subtypes of either component. Comparing these percentage uncertainties with the two other methods of characterisation mentioned -- spectral decomposition \citep{Rebassa-Mansergas2012} and VOSA SED fitting \citep{Rebassa-Mansergas2021} -- demonstrates that using the eclipse gives similar or better precision. For the WD mass, the median percentage uncertainty of the SDSS sample \citep{Rebassa-Mansergas2012} is 18 percent with a 16th percentile (comparable to one standard deviation below the mean for non-normal distributions) of 7 per cent. For the VOSA sample \citep{Rebassa-Mansergas2021} the median is 11 per cent with a 16th percentile of 6 per cent. Our mass determinations are therefore significantly more precise than either of these methods. To illustrate this point further, our least precise measurement of WD mass is more precise than 98.5 per cent of WD mass measurements in the SDSS sample and better than all WD mass measurements in the VOSA sample. For the WD temperature our uncertainties are much more comparable, with a median uncertainty of 3 per cent and a maximum of 5 per cent. This is compared with a median value of 4 per cent for the SDSS sample.

Comparing our best fit parameters to the published, model-independent values (\autoref{tab:old_systems}) additionally demonstrates that a purely photometric approach relying only on eclipse photometry can yield parameters with greater reliability and accuracy than from low-resolution spectroscopy. This improvement on the spectroscopic method is particularly obvious in SDSS~J1028$+$0931 where the contribution from the M~dwarf companion is significant. For all three systems, our parameters are consistent to within $2\sigma$ of the published parameters with most values accurate to better than $\approx5$ per cent. This level of accuracy is better than the original goal of $3\sigma$.

Possible sources of systematic error that we haven't accounted for include the assumption of thick, DA WD models. Although this is consistent with the findings of \citet{Parsons2017a}, it will lead to systematic errors if used for a system containing a WD with a thin hydrogen atmosphere or a helium atmosphere. Additionally, many of the systems considered here contain WDs with masses in the range where theorised hybrid WDs lie \citep{Zenati2019}. Although these have not been confirmed observationally, with a tentative suggestion from \citet{Parsons2020} that has been supported by \citep{Romero2021}, a hybrid core would introduce a similar error into the WD parameters due to an incorrect mass-radius relation. For the secondary, the best-fit effective temperature can be affected by the presence of star spots on the surface or from irradiation effects due to a hot WD, as seems to be the case in NN~Ser. It is also worth mentioning that the statistical uncertainties on the secondary temperature resulting from the MCMC fit are very likely underestimated. The PHOENIX \citep{Husser2013} model grid that we use has a $100~\rm{K}$ resolution in effective temperature and so any uncertainties much below this level are unlikely to represent the true error.

Given the success of eclipse photometry for initial characterisation of WDdM systems, the method will be applicable to LSST data. How well this works will depend on the quality of the absolute flux calibration as well as the final survey strategy, particularly whether the individual 15~s images or photometry are available. This is due to the need to resolve the sharp eclipse features in order to constrain the radii (and hence, masses) of the components. As previously mentioned, many systems discovered by LSST will have little to no parallax information initially (LSST will provide parallax measurements to many of these as the survey progresses). Although fitting the eclipse photometry is still possible in this case, it is more difficult to flag when a fit is converging to erroneous values as a result of systematics.

\section{Conclusions}

We have demonstrated that -- when combined with {\it Gaia} parallax measurements -- high-cadence, multi-colour eclipse photometry can be used to determine masses and temperatures of WDdM binaries more reliably than low-resolution spectroscopy, achieving a precision of better than 5 per cent on the WD parameters and better than 6 per cent for the M~dwarf, making future follow up of these systems easier and more robust. The use of the primary eclipse also guarantees that the photometric SEDs are analysed at the same orbital phase, preventing any possible issues that may arise from using the Virtual Observatory SED Analyser to fit the system.

Additionally, as well as being able to be used for fainter systems than spectroscopic methods, the photometric nature of this method is better equipped to find systems of particular interest such as those displaying variability due to magnetic or pulsating WDs; high-mass WDs, from their sharp eclipse features; or systems with brown dwarf companions that would otherwise be washed out in the optical if not for the clear eclipses. The lack of need for ID spectroscopy also makes this method more time efficient, with the high-cadence photometry being reusable for any high precision follow up work (unlike ID spectroscopy which is often not useful beyond the initial identification).

We have used this method to determine parameters for three new PCEBs, determining two to contain hot, helium-core WDs with low mass companions (one of which is near the brown dwarf transition regime), and one to be a possibly detached CV close to coming back into contact.

\section*{Acknowledgements}

SGP acknowledges the support of the UK's Science and Technology Facilities Council (STFC) Ernest Rutherford Fellowship. IP and TRM acknowledge support from the STFC, grant ST/T000406/1 and a Leverhulme Research Fellowship. VSD, HiPERCAM, and ULTRACAM are supported by the STFC.
Based on observations made with the Gran Telescopio Canarias (GTC), installed in the Spanish Observatorio del Roque de los Muchachos of the Instituto de Astrofísica de Canarias, in the island of La Palma.
This work has made use of data from the European Space Agency (ESA) mission
{\it Gaia} (\url{https://www.cosmos.esa.int/gaia}), processed by the {\it Gaia}
Data Processing and Analysis Consortium (DPAC,
\url{https://www.cosmos.esa.int/web/gaia/dpac/consortium}). Funding for the DPAC
has been provided by national institutions, in particular the institutions
participating in the {\it Gaia} Multilateral Agreement.
For the purpose of open access, the author has applied a Creative Commons Attribution (CC BY) licence to any Author Accepted Manuscript version arising.

\section*{Data Availability}

The data underlying this article will be shared upon reasonable request to the corresponding author.



\bibliographystyle{mnras}
\bibliography{paper} 




\appendix

\section{HiPERCAM photometric system}
\label{section:appendix}

\begin{figure}
    \centering
    \includegraphics[width=\columnwidth]{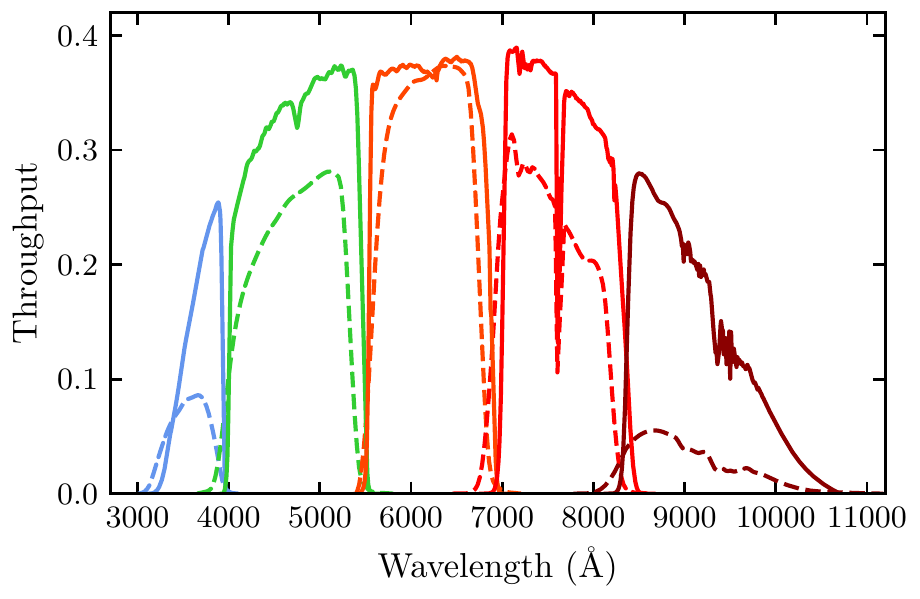}
    \caption{Filter profiles plotted for both HiPERCAM Super-SDSS ($u_{s}g_{s}r_{s}i_{s}z_{s}$) \citep{Dhillon2021} and SDSS ($u'g'r'i'z'$) \citep{Fukugita1996} photometric systems, both including the instrument, telescope, and atmosphere. Solid lines indicate the HiPERCAM system while dashed lines show the SDSS system.}
    \label{fig:filter_profiles}
\end{figure}

As previously mentioned, while the HiPERCAM photometric system approximates the SDSS system (\autoref{fig:filter_profiles}), there are some departures, most significantly in the $u_{s}$ band where the difference is on the order of a few tenths of magnitudes. To assess any colour terms we follow the procedure of \citet{Wild2021}, using synthetic photometry of main~sequence stars and WDs to provide corrections as a function of SDSS colour.

\begin{figure*}
 \includegraphics[width=0.9\textwidth]{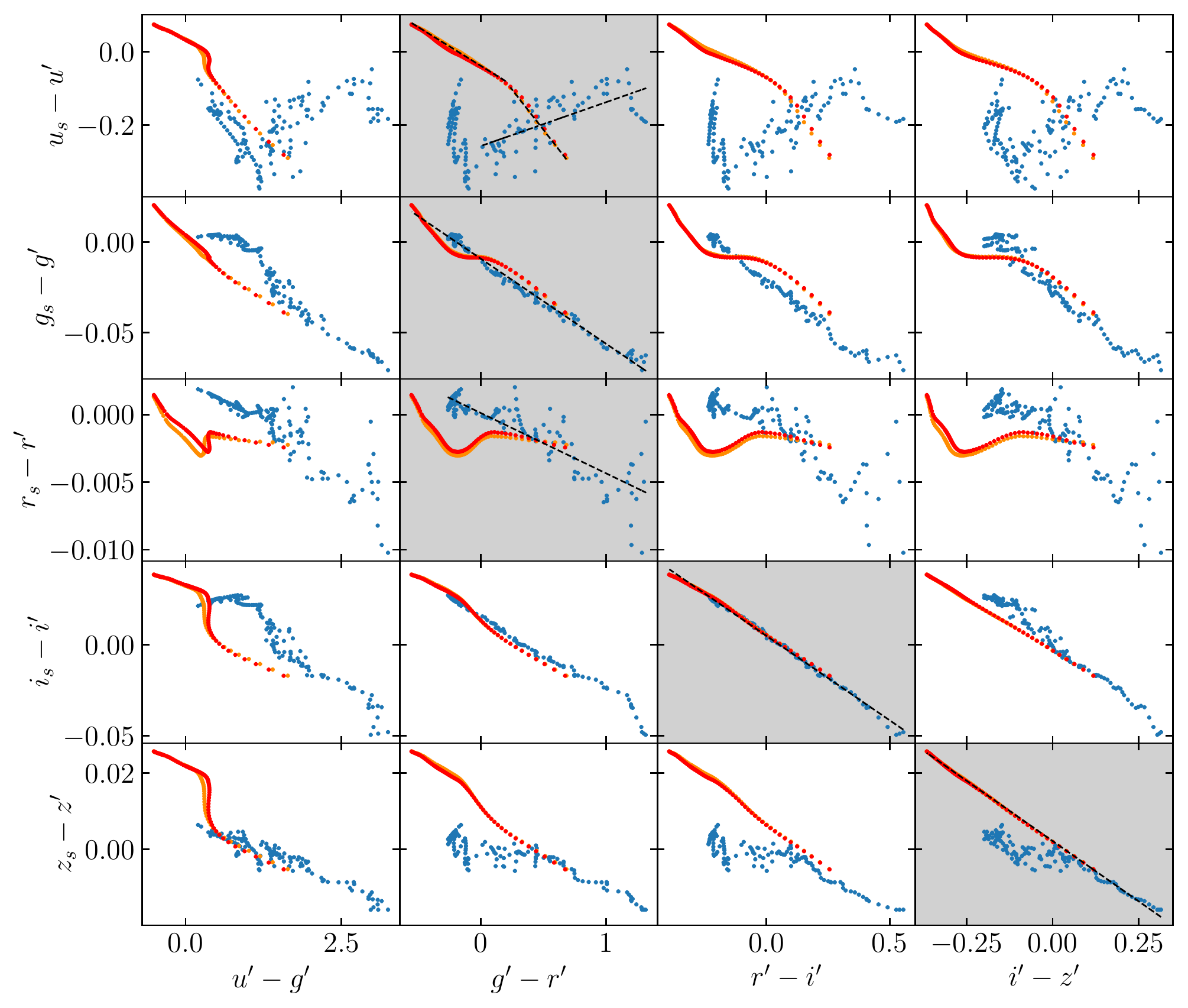}
 \caption{Magnitude offsets between the HiPERCAM Super-SDSS ($u_{s}g_{s}r_{s}i_{s}z_{s}$) photometric system \citep{Dhillon2021} and the SDSS primed ($u'g'r'i'z'$) photometric system \citep{Fukugita1996} as a function of SDSS colour for main~sequence stars \citep{Allard2012} (blue) with $\rm{age}=10^{8.5}~\rm{yr}$ and for WDs \citep{Koester2010} with a $\log(g)=8.0$ (red) and $\log(g)=8.5$ (orange). Shaded plots indicate relations to which colour terms are fit and these best fit corrections (listed in Tables \ref{tab:hcam_corrections_WD2} and \ref{tab:hcam_corrections_MS2}) are shown by a black dashed line.}
 \label{fig:hcam_corrections}
\end{figure*}

For main~sequence stars we use PHOENIX spectral models \citep{Allard2012} at effective temperatures and surface gravities defined by a MIST \citep{Dotter2016, Choi2016} isochrone with an $\rm{age}=10^{8.5}~\rm{yr}$, covering masses ranging from $0.1-3~\rm{M_{\odot}}$ and surface gravities from $3.7-4.7$. For WDs we use \citet{Koester2010} models with a $\log(g~\rm{[cgs]})=8.5$ over the full range of available model temperatures. We then use \autoref{eqn:syn_phot} to generate synthetic photometry for the HiPERCAM Super-SDSS and SDSS systems. Corrections as a function of colour are shown in \autoref{fig:hcam_corrections} with best fit colour terms shown in Tables \ref{tab:hcam_corrections_WD2} and \ref{tab:hcam_corrections_MS2} for WDs and main~sequence stars respectively. Corrections should be possible to an accuracy of a couple of per cent in the $g_{s}, r_{s}, i_{s}$, and $z_{s}$ bands using main~sequence stars but it is clear that no easy correction can be made for $u_{s}$ where there is no consistent correlation with colour. As such, any correction to the $u_s$ band magnitudes using main~sequence stars in combination with these colour terms should be avoided if possible. Using WDs make this a lot easier with tight relations between the colour and corrections that only weakly depend on the surface gravity of the WD (as demonstrated by the similarity of $\log(g)=8.0$ and $\log(g)=8.5$ relations).

In order to make flux calibration simpler and more robust in future, we define a set of HiPERCAM standard stars (\autoref{tab:hcam_mags}). These standard stars are the {\it Gaia} spectro-photometric standard stars collated and evaluated by \citet{Pancino2012, Altavilla2015, Marinoni2016, Altavilla2021, Pancino2021} who provide high quality spectra of these standards covering the full range of the HiPERCAM photometric system. We again use \autoref{eqn:syn_phot} to produce synthetic AB magnitudes for these stars in both the HiPERCAM and ULTRACAM systems. \citet{Pancino2021} mentions that the scatter in the spectro-photometric standard stars when compared with literature is of order $1$ per cent with discrepant behavior of a similar order in the red for faint blue stars. Additionally, due to the use of spectral models to extend the flux tables below $400~\rm{nm}$ and above $800~\rm{nm}$, we estimate an uncertainty of $2$ per cent in $u_{s}$ and $z_{s}$, and $1$ per cent in $g_{s}$, $r_{s}$, and $i_{s}$.

\begin{table}
    \centering
    \caption{SDSS to HiPERCAM Super-SDSS colour terms for WDs. Validity shows the range of colours spanned by the models that the colour terms were fit to. These take the form of a straight line, e.g. $u_{s}-u'=-0.211(g'-r')-0.038$}
    \label{tab:hcam_corrections_WD2}
    \begin{tabular}{ccccc}
        \hline
        Correction & Gradient & Variable & $y$-intercept & Validity \\
        \hline
        $u_s-u'$ & $-0.211$ & $g'-r'$ & $-0.038$ & $-0.55<g'-r'\leq0.20$ \\
        & $-0.438$ & $g'-r'$ & $0.006$ & $0.20<g'-r'\leq0.70$ \\
        $g_s-g'$ & $-0.047$ & $g'-r'$ & $-0.009$ & $-0.55<g'-r'\leq0.70$ \\
        $r_s-r'$ & -- & -- & -- & -- \\
        $i_s-i'$ & $-0.093$ & $r'-i'$ & $0.005$ & $-0.40<r'-i'\leq0.25$ \\
        $z_s-z'$ & $-0.047$ & $i'-z'$ & $-0.009$ & $-0.35<i'-z'\leq0.10$ \\
        \hline
    \end{tabular}
\end{table}

\begin{table}
    \centering
    \caption{As \autoref{tab:hcam_corrections_WD2} but for main~sequence models.}
    \label{tab:hcam_corrections_MS2}
    \begin{tabular}{ccccc}
        \hline
        Correction & Gradient & Variable & $y$-intercept & Validity \\
        \hline
        $u_s-u'$ & $0.120$ & $g'-r'$ & $-0.257$ & $0.00<g'-r'\leq1.30$ \\
        $g_s-g'$ & $-0.047$ & $g'-r'$ & $-0.009$ & $-0.25<g'-r'\leq1.30$ \\
        $r_s-r'$ & -0.004 & $g'-r'$ & 0.000 & $-0.25<g'-r'\leq1.30$ \\
        $i_s-i'$ & $-0.093$ & $r'-i'$ & $0.005$ & $-0.25<g'-r'\leq0.55$ \\
        $z_s-z'$ & $0.000$ & $i'-z'$ & $0.000$ & $-0.20<i'-z'\leq0.10$ \\
        & $-0.047$ & $i'-z'$ & $-0.009$ & $0.10<i'-z'\leq0.30$ \\
        \hline
    \end{tabular}
\end{table}

\begin{landscape}
\begin{table}
\caption{{\it Gaia} spectro-photometric standard stars \citep{Pancino2012, Altavilla2015, Marinoni2016, Altavilla2021, Pancino2021} with AB magnitudes computed for the HiPERCAM and ULTRACAM Super-SDSS ($u_{s}g_{s}r_{s}i_{s}z_{s}$) photometric systems using the flux tables of \citet{Pancino2021}. The 'Type' column indicates the status of a flux standard as either 'Pillar', 'Primary', or 'Secondary' as described in \citet{Pancino2012} ('0', '1', and '2' respectively in the table). The pillars denoted here are the same three stars on which the CALSPEC system is based \citep{Bohlin1995} and the primary stars are all bright, well-known spectro-photometric standards that are already tied to -- or are easy to tie to -- the CALSPEC flux scale. Secondary standards are then calibrated from these primary stars. The 'Stability' column shows which standards have been confirmed as photometrically constant by the variability monitoring campaign \citep{Marinoni2016}. Standards that are not yet confirmed as photometrically constant are still considered likely to be constant \citep[see section 3.4]{Marinoni2016}. We therefore choose not to discard them.}
\label{tab:hcam_mags}
\begin{tabular}{@{\extracolsep{4pt}}lccccccccccccccc@{}}
\hline
\hline
& & & \multicolumn{5}{c}{HiPERCAM} & \multicolumn{5}{c}{ULTRACAM} & & \\
\cline{4-8}
\cline{9-13}
Name & RA (J2000) & DEC (J2000) & $u_{s}$ & $g_{s}$ & $r_{s}$ & $i_{s}$ & $z_{s}$ & $u_{s}$ & $g_{s}$ & $r_{s}$ & $i_{s}$ & $z_{s}$ & Type & SpType & Stability \\
\hline
WD~0004+330 & 00:07:32.26 & +33:17:27.60 & 13.158 & 13.531 & 14.023 & 14.426 & 14.742 & 13.125 & 13.523 & 14.042 & 14.424 & 14.781 & 2 & DA1 &  \\
WD~0018$-$267 & 00:21:30.73 & $-$26:26:11.46 & 15.252 & 14.083 & 13.621 & 13.431 & 13.391 & 15.302 & 14.094 & 13.610 & 13.433 & 13.385 & 2 & DA9 &  \\
WD~0038+555 & 00:41:21.99 & +55:50:08.40 & 14.014 & 13.980 & 14.106 & 14.280 & 14.442 & 14.019 & 13.980 & 14.113 & 14.280 & 14.471 & 2 & DQ5 & Confirmed \\
LTT~377 & 00:41:30.47 & $-$33:37:32.03 & 13.783 & 11.279 & 9.973 & 9.275 & 9.060 & 13.889 & 11.296 & 9.948 & 9.283 & 9.062 & 2 & K9 & Confirmed \\
WD~0046+051 & 00:49:09.90 & +05:23:19.01 & 13.578 & 12.556 & 12.301 & 12.268 & 12.395 & 13.574 & 12.567 & 12.298 & 12.268 & 12.407 & 2 & DZ7 & Confirmed \\
WD~0047$-$524 & 00:50:03.68 & $-$52:08:15.60 & 14.172 & 14.059 & 14.422 & 14.741 & 15.036 & 14.171 & 14.054 & 14.439 & 14.738 & 15.069 & 2 & DA2 &  \\
WD~0050$-$332 & 00:53:17.44 & $-$32:59:56.60 & 12.767 & 13.095 & 13.589 & 13.934 & 14.191 & 12.742 & 13.086 & 13.607 & 13.932 & 14.222 & 2 & DA1 &  \\
WD~0109$-$264 & 01:12:11.65 & $-$26:13:27.69 & 12.695 & 12.883 & 13.355 & 13.724 & 14.025 & 12.691 & 12.875 & 13.372 & 13.722 & 14.058 & 2 & DA1 &  \\
WD~0123$-$262 & 01:25:24.45 & $-$26:00:43.90 & 15.531 & 15.144 & 15.008 & 15.013 & 15.130 & 15.580 & 15.149 & 15.006 & 15.013 & 15.145 & 2 & DC & Confirmed \\
G245-31 & 01:38:39.39 & +69:38:01.50 & 15.886 & 14.849 & 14.243 & 13.983 & 13.865 & 15.953 & 14.860 & 14.228 & 13.985 & 13.865 & 2 & K & Confirmed \\
GJ~70 & 01:43:20.18 & +04:19:17.97 & 13.931 & 11.815 & 10.445 & 9.286 & 8.799 & 14.173 & 11.834 & 10.416 & 9.297 & 8.761 & 2 & M2 & Confirmed \\
LTT~1020 & 01:54:50.27 & $-$27:28:35.74 & 12.542 & 11.736 & 11.361 & 11.226 & 11.215 & 12.614 & 11.745 & 11.353 & 11.227 & 11.216 & 1 & G & Confirmed \\
LP~885-23 & 02:07:06.33 & $-$30:24:22.90 & 17.474 & 14.894 & 13.549 & 12.138 & 11.558 & 17.620 & 14.914 & 13.515 & 12.152 & 11.516 & 2 & M0 & Confirmed \\
EGGR~21 & 03:10:31.02 & $-$68:36:03.39 & 11.468 & 11.258 & 11.563 & 11.855 & 12.151 & 11.467 & 11.254 & 11.578 & 11.851 & 12.179 & 1 & DA3 & Confirmed \\
HG~7-15 & 03:48:11.86 & 07:08:46.47 & 14.059 & 11.513 & 10.274 & 9.710 & 9.426 & 14.134 & 11.528 & 10.250 & 9.716 & 9.409 & 2 & M1 & Confirmed \\
LTT~1788 & 03:48:22.67 & $-$39:08:37.20 & 14.013 & 13.327 & 13.025 & 12.907 & 12.902 & 14.098 & 13.334 & 13.018 & 12.909 & 12.908 & 1 & F & Confirmed \\
GD~50 & 03:48:50.20 & $-$00:58:31.20 & 13.401 & 13.797 & 14.290 & 14.683 & 15.040 & 13.367 & 13.788 & 14.308 & 14.681 & 15.081 & 1 & DA2 &  \\
HZ~2 & 04:12:43.55 & +11:51:49.00 & 13.719 & 13.709 & 14.076 & 14.410 & 14.688 & 13.706 & 13.703 & 14.092 & 14.407 & 14.714 & 1 & DA3 &  \\
WD~0455$-$282 & 04:57:13.90 & $-$28:07:54.00 & 13.262 & 13.649 & 14.181 & 14.581 & 14.887 & 13.237 & 13.640 & 14.199 & 14.578 & 14.930 & 2 & DA1 &  \\
WD~0501$-$289 & 05:03:55.51 & $-$28:54:34.57 & 13.086 & 13.583 & 14.122 & 14.559 & 14.899 & 13.057 & 13.572 & 14.141 & 14.557 & 14.936 & 2 & DO & Confirmed \\
G191-B2B & 05:05:30.61 & +52:49:51.95 & 11.069 & 11.492 & 12.010 & 12.427 & 12.778 & 11.031 & 11.483 & 12.029 & 12.423 & 12.813 & 0 & DA0 & Confirmed \\
GD~71 & 05:52:27.63 & +15:53:13.37 & 12.480 & 12.783 & 13.258 & 13.654 & 14.000 & 12.451 & 12.775 & 13.277 & 13.651 & 14.034 & 0 & DA1 & Confirmed \\
LTT~2415 & 05:56:24.74 & $-$27:51:32.35 & 13.045 & 12.348 & 12.115 & 12.032 & 12.041 & 13.156 & 12.354 & 12.110 & 12.033 & 12.053 & 1 & G & Confirmed \\
HD~270477 & 05:59:33.36 & $-$67:01:13.72 & 11.392 & 10.501 & 10.316 & 10.306 & 10.354 & 11.528 & 10.506 & 10.314 & 10.306 & 10.363 & 2 & F8 &  \\
HD~271747 & 05:59:58.62 & $-$66:06:08.91 & 12.824 & 11.689 & 11.303 & 11.194 & 11.196 & 12.904 & 11.699 & 11.294 & 11.195 & 11.200 & 2 & G0 &  \\
HD~271759 & 06:00:41.34 & $-$66:03:14.03 & 11.803 & 10.862 & 10.877 & 10.991 & 11.090 & 12.011 & 10.863 & 10.881 & 10.990 & 11.100 & 2 & A2 & Confirmed \\
HD~271783 & 06:02:11.36 & $-$66:34:59.13 & 13.209 & 12.148 & 11.783 & 11.697 & 11.755 & 13.310 & 12.157 & 11.775 & 11.698 & 11.761 & 2 & F5 &  \\
WD~0604$-$203 & 06:06:13.39 & $-$20:21:07.20 & 11.773 & 11.831 & 12.269 & 12.608 & 12.903 & 11.788 & 11.825 & 12.285 & 12.605 & 12.937 & 2 & DA & Confirmed \\
WD~0621$-$376 & 06:23:12.63 & $-$37:41:28.01 & 11.372 & 11.792 & 12.325 & 12.741 & 13.069 & 11.341 & 11.783 & 12.343 & 12.738 & 13.111 & 2 & DA1 & Confirmed \\
HILT~600 & 06:45:13.37 & +02:08:14.70 & 10.751 & 10.458 & 10.452 & 10.535 & 10.631 & 10.820 & 10.458 & 10.455 & 10.535 & 10.651 & 1 & B1 & Confirmed \\
WD~0644+375 & 06:47:37.99 & +37:30:57.07 & 11.832 & 11.882 & 12.258 & 12.606 & 12.906 & 11.814 & 11.876 & 12.275 & 12.603 & 12.944 & 2 & DA2 &  \\
WD~0646$-$253 & 06:48:56.09 & $-$25:23:47.00 & 13.220 & 13.419 & 13.863 & 14.247 & 14.507 & 13.199 & 13.411 & 13.881 & 14.244 & 14.532 & 2 & DA2 & Confirmed \\
G193-26 & 07:03:26.29 & +54:52:06.00 & 13.955 & 13.223 & 12.812 & 12.635 & 12.567 & 14.019 & 13.232 & 12.802 & 12.637 & 12.575 & 2 & G & Confirmed \\
WD~0721$-$276 & 07:23:20.10 & $-$27:47:21.60 & 13.973 & 14.297 & 14.771 & 15.149 & 15.470 & 13.949 & 14.288 & 14.788 & 15.145 & 15.511 & 2 & DA1 & Confirmed \\
LTT~3218 & 08:41:32.56 & $-$32:56:34.90 & 12.248 & 11.875 & 11.903 & 12.025 & 12.186 & 12.274 & 11.876 & 11.908 & 12.023 & 12.203 & 1 & DA & Confirmed \\
G114-25 & 08:59:03.37 & $-$06:23:46.19 & 12.976 & 12.169 & 11.741 & 11.551 & 11.519 & 13.050 & 12.179 & 11.730 & 11.552 & 11.524 & 2 & F7 & Confirmed \\
WD~0859$-$039 & 09:02:17.30 & $-$04:06:55.45 & 12.859 & 12.986 & 13.385 & 13.739 & 14.055 & 12.841 & 12.979 & 13.401 & 13.736 & 14.096 & 2 & DA2 & Confirmed \\
GD~108 & 10:00:47.37 & $-$07:33:30.50 & 13.195 & 13.337 & 13.756 & 14.101 & 14.393 & 13.198 & 13.328 & 13.772 & 14.099 & 14.431 & 1 & B & Confirmed \\
LTT~3864 & 10:32:13.60 & $-$35:37:41.80 & 13.140 & 12.358 & 12.034 & 11.914 & 11.902 & 13.229 & 12.366 & 12.027 & 11.915 & 11.898 & 1 & F & Confirmed \\
WD~1031$-$114 & 10:33:42.76 & $-$11:41:38.35 & 12.638 & 12.808 & 13.220 & 13.569 & 13.871 & 12.619 & 12.801 & 13.236 & 13.566 & 13.910 & 2 & DA2 & Confirmed \\
WD~1034+001 & 10:37:03.81 & $-$00:08:19.30 & 12.451 & 12.925 & 13.470 & 13.904 & 14.246 & 12.416 & 12.916 & 13.490 & 13.901 & 14.288 & 2 & DOZ1 & Confirmed \\
Feige~34 & 10:39:36.74 & +43:06:09.25 & 10.451 & 10.880 & 11.393 & 11.781 & 12.108 & 10.420 & 10.871 & 11.411 & 11.777 & 12.152 & 1 & DO & Confirmed \\
WD~1105$-$048 & 11:07:59.95 & $-$05:09:25.90 & 13.197 & 12.957 & 13.245 & 13.544 & 13.814 & 13.193 & 12.953 & 13.260 & 13.541 & 13.850 & 2 & DA3 & Confirmed \\
SDSS~J1138+5729 & 11:38:02.62 & +57:29:23.89 & 15.855 & 15.089 & 14.989 & 15.010 & 15.068 & 16.003 & 15.093 & 14.989 & 15.011 & 15.082 & 2 & A0/F3 & Confirmed \\

\end{tabular}
\end{table}
\end{landscape}

\begin{landscape}
\begin{table}
\contcaption{}
\label{tab:continued}
\begin{tabular}{@{\extracolsep{4pt}}lccccccccccccccc@{}}
\hline
\hline
& & & \multicolumn{5}{c}{HiPERCAM} & \multicolumn{5}{c}{ULTRACAM} & & \\
\cline{4-8}
\cline{9-13}
Name & RA (J2000) & DEC (J2000) & $u_{s}$ & $g_{s}$ & $r_{s}$ & $i_{s}$ & $z_{s}$ & $u_{s}$ & $g_{s}$ & $r_{s}$ & $i_{s}$ & $z_{s}$ & Type & SpType & Stability \\
\hline
LTT~4364 & 11:45:42.92 & $-$64:50:29.46 & 11.745 & 11.516 & 11.477 & 11.557 & 11.688 & 11.778 & 11.518 & 11.479 & 11.556 & 11.709 & 1 & DQ6 & Confirmed \\
Feige~56 & 12:06:47.23 & +11:40:12.64 & 11.068 & 10.872 & 11.193 & 11.496 & 11.713 & 11.179 & 10.867 & 11.207 & 11.494 & 11.741 & 1 & B5p & Confirmed \\
HD~106355 & 12:14:10.53 & $-$17:14:20.19 & 12.822 & 10.641 & 9.738 & 9.355 & 9.146 & 12.858 & 10.660 & 9.716 & 9.358 & 9.119 & 2 & G8IV & Confirmed \\
Feige~66 & 12:37:23.52 & +25:03:59.87 & 9.924 & 10.236 & 10.731 & 11.138 & 11.464 & 9.901 & 10.228 & 10.750 & 11.134 & 11.505 & 1 & O & Confirmed \\
SA~104-428 & 12:41:41.28 & $-$00:26:26.20 & 15.035 & 13.088 & 12.342 & 12.061 & 11.914 & 15.087 & 13.105 & 12.324 & 12.063 & 11.911 & 2 & G8 & Confirmed \\
SA~104-490 & 12:44:33.46 & $-$00:25:51.70 & 13.799 & 12.779 & 12.448 & 12.366 & 12.367 & 13.902 & 12.788 & 12.441 & 12.367 & 12.378 & 2 & G3 & Confirmed \\
GD~153 & 12:57:02.33 & +22:01:52.52 & 12.723 & 13.081 & 13.575 & 13.986 & 14.326 & 12.692 & 13.072 & 13.594 & 13.984 & 14.361 & 0 & DA1 & Confirmed \\
G14-24 & 13:02:01.58 & $-$02:05:21.42 & 14.208 & 13.148 & 12.584 & 12.337 & 12.224 & 14.275 & 13.160 & 12.570 & 12.339 & 12.226 & 2 & K0 & Confirmed \\
GJ~2097 & 13:07:04.31 & +20:48:38.54 & 15.809 & 13.219 & 11.850 & 10.787 & 10.313 & 15.946 & 13.237 & 11.822 & 10.797 & 10.289 & 2 & M1 & Confirmed \\
HZ~44 & 13:23:35.26 & +36:07:59.51 & 10.999 & 11.389 & 11.885 & 12.310 & 12.653 & 10.971 & 11.382 & 11.904 & 12.307 & 12.685 & 1 & O & Confirmed \\
SA~105-663 & 13:37:30.34 & $-$00:13:17.37 & 9.755 & 8.856 & 8.730 & 8.706 & 8.795 & 9.929 & 8.861 & 8.729 & 8.708 & 8.799 & 1 & F2 & Confirmed \\
GJ~521 & 13:39:24.10 & +46:11:11.37 & 13.354 & 10.933 & 9.673 & 8.743 & 8.320 & 13.520 & 10.948 & 9.645 & 8.752 & 8.301 & 2 & M2 & Confirmed \\
HD~121968 & 13:58:51.17 & $-$02:54:52.32 & 9.924 & 10.041 & 10.423 & 10.754 & 11.023 & 9.950 & 10.034 & 10.437 & 10.752 & 11.052 & 1 & B1 & Confirmed \\
WD~1408+323 & 14:10:26.95 & +32:08:36.10 & 13.924 & 13.833 & 14.169 & 14.493 & 14.790 & 13.912 & 13.828 & 14.185 & 14.491 & 14.827 & 2 & DA3 &  \\
SDSS~J1429+3928 & 14:29:51.06 & +39:28:25.43 & 15.865 & 15.006 & 15.042 & 15.150 & 15.233 & 16.082 & 15.007 & 15.047 & 15.151 & 15.247 & 2 & A0 & Confirmed \\
G15-10 & 15:09:46.02 & $-$04:45:06.61 & 13.170 & 12.310 & 11.816 & 11.608 & 11.548 & 13.232 & 12.320 & 11.803 & 11.610 & 11.548 & 2 & G2 & Confirmed \\
WD~1509+322 & 15:11:27.66 & +32:04:17.80 & 14.267 & 14.009 & 14.276 & 14.567 & 14.825 & 14.265 & 14.005 & 14.292 & 14.565 & 14.860 & 2 & DA3 & Confirmed \\
2MASS~J1517+0202 & 15:17:38.64 & +02:02:25.60 & 16.234 & 14.558 & 13.813 & 13.474 & 13.313 & 16.328 & 14.575 & 13.795 & 13.478 & 13.308 & 2 &  & Confirmed \\
G167-50 & 15:35:31.55 & +27:51:02.20 & 14.780 & 13.824 & 13.305 & 13.090 & 13.014 & 14.829 & 13.834 & 13.294 & 13.092 & 13.016 & 2 & G & Confirmed \\
SA~107-544 & 15:36:48.10 & $-$00:15:07.11 & 10.163 & 9.152 & 8.973 & 8.945 & 8.987 & 10.339 & 9.157 & 8.970 & 8.946 & 8.986 & 1 & F3 & Confirmed \\
LTT~6248 & 15:38:59.66 & $-$28:35:36.87 & 12.720 & 11.993 & 11.659 & 11.526 & 11.512 & 12.817 & 11.999 & 11.651 & 11.527 & 11.523 & 1 & A & Confirmed \\
G179-54 & 15:46:08.25 & +39:14:16.40 & 14.612 & 13.761 & 13.322 & 13.148 & 13.088 & 14.668 & 13.771 & 13.312 & 13.150 & 13.094 & 2 & F & Confirmed \\
G224-83 & 15:46:14.68 & +62:26:39.60 & 13.777 & 12.984 & 12.563 & 12.391 & 12.333 & 13.838 & 12.993 & 12.553 & 12.393 & 12.339 & 2 & K & Confirmed \\
G16-20 & 15:58:18.62 & +02:03:06.11 & 11.966 & 11.042 & 10.615 & 10.441 & 10.373 & 12.069 & 11.051 & 10.605 & 10.442 & 10.380 & 2 & K & Confirmed \\
GSPC~P177-D & 15:59:13.57 & +47:36:41.90 & 14.936 & 13.723 & 13.279 & 13.133 & 13.116 & 14.991 & 13.734 & 13.270 & 13.135 & 13.113 & 1 & G0 & Confirmed \\
WD~1606+422 & 16:08:22.20 & +42:05:43.20 & 14.163 & 13.758 & 13.959 & 14.201 & 14.435 & 14.173 & 13.756 & 13.972 & 14.200 & 14.465 & 2 & DA4 & Confirmed \\
WD~1615$-$154 & 16:17:55.26 & $-$15:35:51.90 & 12.919 & 13.198 & 13.649 & 14.002 & 14.285 & 12.896 & 13.189 & 13.666 & 13.999 & 14.329 & 2 & DA2 &  \\
EGGR~274 & 16:23:33.84 & $-$39:13:46.16 & 10.712 & 10.810 & 11.237 & 11.583 & 11.883 & 10.699 & 10.804 & 11.255 & 11.580 & 11.922 & 1 & DA2 &  \\
G180-58 & 16:28:16.87 & +44:40:38.28 & 12.554 & 11.601 & 11.096 & 10.889 & 10.804 & 12.611 & 11.612 & 11.085 & 10.891 & 10.808 & 2 & G/K & Confirmed \\
WD~1626+368 & 16:28:25.03 & +36:46:15.40 & 14.028 & 13.805 & 13.856 & 13.995 & 14.168 & 13.993 & 13.808 & 13.862 & 13.994 & 14.193 & 2 & DZA6 & Confirmed \\
G170-47 & 17:32:41.63 & +23:44:11.64 & 10.090 & 9.191 & 8.728 & 8.529 & 8.457 & 10.198 & 9.201 & 8.717 & 8.530 & 8.459 & 2 & G0 & Confirmed \\
2MASS~J17430+6655 & 17:43:04.48 & +66:55:01.60 & 14.495 & 13.558 & 13.531 & 13.605 & 13.665 & 14.698 & 13.560 & 13.535 & 13.605 & 13.670 & 1 & A5 & Confirmed \\
RMC2005~KF08T3 & 17:55:16.23 & +66:10:11.70 & 15.568 & 13.731 & 13.006 & 12.738 & 12.605 & 15.601 & 13.747 & 12.989 & 12.740 & 12.594 & 1 & K0 & Confirmed \\
TYC~4212-455-1 & 17:57:13.25 & +67:03:40.90 & 12.811 & 11.842 & 11.816 & 11.905 & 11.970 & 13.022 & 11.845 & 11.819 & 11.905 & 11.979 & 2 & A3 & Confirmed \\
RMC2005~KF06T2 & 17:58:37.99 & +66:46:52.20 & 16.856 & 14.518 & 13.602 & 13.233 & 13.038 & 16.930 & 14.538 & 13.579 & 13.237 & 13.020 & 1 & K1 & Confirmed \\
BD+66~1071 & 18:02:10.92 & +66:12:26.39 & 11.579 & 10.688 & 10.475 & 10.449 & 10.479 & 11.700 & 10.694 & 10.472 & 10.450 & 10.489 & 2 & F5 & Confirmed \\
TYC~4209-1396-1 & 18:05:29.28 & +64:27:52.00 & 13.127 & 12.238 & 12.372 & 12.536 & 12.649 & 13.358 & 12.236 & 12.380 & 12.535 & 12.654 & 1 & A6 & Confirmed \\
TYC~4205-1677-1 & 18:12:09.57 & +63:29:42.30 & 12.715 & 11.731 & 11.781 & 11.911 & 12.005 & 12.929 & 11.733 & 11.787 & 11.911 & 12.010 & 1 & A5 & Confirmed \\
LTT~7379 & 18:36:25.95 & $-$44:18:36.94 & 11.488 & 10.472 & 10.077 & 9.924 & 9.943 & 11.560 & 10.482 & 10.069 & 9.925 & 9.955 & 1 & G0 & Confirmed \\
G184-17 & 18:40:29.27 & +19:36:06.65 & 15.621 & 14.453 & 13.845 & 13.607 & 13.490 & 15.661 & 14.466 & 13.832 & 13.609 & 13.482 & 2 & K & Confirmed \\
WD~1845+019 & 18:47:39.08 & +01:57:35.62 & 12.513 & 12.726 & 13.156 & 13.396 & 13.418 & 12.489 & 12.719 & 13.173 & 13.395 & 13.426 & 2 & DA2 &  \\
GJ~745A & 19:07:05.56 & +20:53:16.97 & 14.176 & 11.573 & 10.171 & 9.093 & 8.575 & 14.315 & 11.591 & 10.145 & 9.104 & 8.544 & 2 & M2 & Confirmed \\
GJ~745B & 19:07:13.20 & +20:52:37.24 & 14.177 & 11.570 & 10.173 & 9.080 & 8.562 & 14.311 & 11.589 & 10.146 & 9.092 & 8.531 & 2 & M2 & Confirmed \\
WD~1914$-$598 & 19:18:44.84 & $-$59:46:33.80 & 14.292 & 14.219 & 14.582 & 14.888 & 15.203 & 14.284 & 14.213 & 14.598 & 14.885 & 15.235 & 2 & DA & Confirmed \\
EGGR~131 & 19:20:34.93 & $-$07:40:00.05 & 12.228 & 12.219 & 12.359 & 12.544 & 12.750 & 12.236 & 12.218 & 12.366 & 12.544 & 12.779 & 1 & DBQA5 & Confirmed \\
WD~1919+145 & 19:21:40.40 & +14:40:43.00 & 13.128 & 12.881 & 13.159 & 13.469 & 13.753 & 13.123 & 12.877 & 13.175 & 13.467 & 13.782 & 2 & DA3 & Confirmed \\
G23-14 & 19:51:49.61 & +05:36:45.84 & 12.159 & 11.059 & 10.500 & 10.268 & 10.164 & 12.242 & 11.071 & 10.487 & 10.270 & 10.161 & 2 & G5 & Confirmed \\
WD~2004$-$605 & 20:09:05.24 & $-$60:25:41.60 & 12.721 & 13.086 & 13.594 & 13.972 & 14.326 & 12.695 & 13.077 & 13.612 & 13.970 & 14.363 & 2 & DA1 & Confirmed \\
LTT~7987 & 20:10:56.85 & $-$30:13:06.64 & 12.390 & 12.126 & 12.420 & 12.699 & 12.979 & 12.389 & 12.122 & 12.435 & 12.696 & 13.011 & 1 & DA4 & Confirmed \\

\end{tabular}
\end{table}
\end{landscape}

\begin{landscape}
\begin{table}
\contcaption{}
\label{tab:continued}
\begin{tabular}{@{\extracolsep{4pt}}lccccccccccccccc@{}}
\hline
\hline
& & & \multicolumn{5}{c}{HiPERCAM} & \multicolumn{5}{c}{ULTRACAM} & & \\
\cline{4-8}
\cline{9-13}
Name & RA (J2000) & DEC (J2000) & $u_{s}$ & $g_{s}$ & $r_{s}$ & $i_{s}$ & $z_{s}$ & $u_{s}$ & $g_{s}$ & $r_{s}$ & $i_{s}$ & $z_{s}$ & Type & SpType & Stability \\
\hline
WD~2032+248 & 20:34:21.88 & +25:03:49.72 & 11.348 & 11.346 & 11.721 & 12.078 & 12.388 & 11.332 & 11.340 & 11.739 & 12.075 & 12.421 & 2 & DA2 &  \\
WD~2034$-$532 & 20:38:16.84 & $-$53:04:25.40 & 14.089 & 14.214 & 14.522 & 14.783 & 15.018 & 14.083 & 14.212 & 14.535 & 14.781 & 15.045 & 2 & DB4 & Confirmed \\
G24-25 & 20:40:16.10 & +00:33:19.74 & 11.677 & 10.806 & 10.406 & 10.266 & 10.216 & 11.756 & 10.815 & 10.398 & 10.268 & 10.212 & 2 & G0 & Confirmed \\
WD~2039$-$682 & 20:44:21.47 & $-$68:05:21.30 & 13.354 & 13.202 & 13.496 & 13.776 & 14.071 & 13.342 & 13.198 & 13.511 & 13.773 & 14.101 & 2 & DA3 & Confirmed \\
WD~2047+372 & 20:49:06.69 & +37:28:13.90 & 13.221 & 12.918 & 13.160 & 13.430 & 13.693 & 13.221 & 12.915 & 13.175 & 13.428 & 13.726 & 2 & DA3 & Confirmed \\
WD~2111+498 & 21:12:44.05 & +50:06:17.80 & 12.463 & 12.792 & 13.270 & 13.662 & 13.994 & 12.435 & 12.783 & 13.289 & 13.659 & 14.030 & 2 & DA1 &  \\
WD~2105$-$820 & 21:13:13.90 & $-$81:49:04.00 & 13.999 & 13.660 & 13.734 & 13.883 & 14.061 & 14.004 & 13.659 & 13.741 & 13.880 & 14.084 & 2 & DA5 & Confirmed \\
WD~2117+539 & 21:18:56.27 & +54:12:41.25 & 12.530 & 12.249 & 12.526 & 12.828 & 13.097 & 12.532 & 12.245 & 12.541 & 12.825 & 13.123 & 2 & DA3 & Confirmed \\
WD~2134+218 & 21:36:36.30 & +22:04:33.00 & 14.454 & 14.329 & 14.650 & 14.966 & 15.232 & 14.442 & 14.324 & 14.666 & 14.964 & 15.265 & 2 & DA3 &  \\
WD~2140+207 & 21:42:41.00 & +20:59:58.24 & 13.425 & 13.245 & 13.250 & 13.354 & 13.480 & 13.445 & 13.247 & 13.253 & 13.354 & 13.499 & 2 & DQ6 & Confirmed \\
G93-48 & 21:52:25.38 & +02:23:19.56 & 12.749 & 12.622 & 12.950 & 13.259 & 13.552 & 12.741 & 12.618 & 12.966 & 13.258 & 13.589 & 1 & DA3 &  \\
WD~2216$-$657 & 22:19:48.35 & $-$65:29:18.11 & 14.633 & 14.484 & 14.589 & 14.742 & 14.920 & 14.606 & 14.485 & 14.596 & 14.740 & 14.940 & 2 & DZ5 & Confirmed \\
LTT~9239 & 22:52:41.03 & $-$20:35:32.89 & 13.228 & 12.325 & 11.905 & 11.725 & 11.729 & 13.292 & 12.335 & 11.895 & 11.726 & 11.734 & 1 & F & Confirmed \\
WD~2251$-$634 & 22:55:10.00 & $-$63:10:27.00 & 14.046 & 13.968 & 14.328 & 14.614 & 14.901 & 14.041 & 13.963 & 14.343 & 14.612 & 14.938 & 2 & DA & Confirmed \\
WD~2309+105 & 23:12:21.62 & +10:47:04.25 & 12.395 & 12.812 & 13.323 & 13.733 & 14.079 & 12.356 & 12.803 & 13.342 & 13.731 & 14.122 & 2 & DA1 &  \\
Feige~110 & 23:19:58.40 & $-$05:09:56.21 & 11.168 & 11.547 & 12.059 & 12.446 & 12.787 & 11.140 & 11.539 & 12.077 & 12.443 & 12.821 & 1 & O & Confirmed \\
WD~2329+407 & 23:31:35.65 & +41:01:30.70 & 13.953 & 13.744 & 14.040 & 14.347 & 14.623 & 13.948 & 13.739 & 14.056 & 14.345 & 14.657 & 2 & DA3 &  \\
WD~2331$-$475 & 23:34:02.20 & $-$47:14:26.50 & 12.760 & 13.149 & 13.662 & 14.042 & 14.317 & 12.730 & 13.140 & 13.680 & 14.039 & 14.345 & 2 & DA1 &  \\
WD~2352+401 & 23:54:56.25 & +40:27:30.10 & 15.229 & 14.990 & 14.917 & 14.977 & 15.110 & 15.260 & 14.992 & 14.917 & 14.977 & 15.137 & 2 & DQ6 & Confirmed \\
\hline
\end{tabular}
\end{table}
\end{landscape}

\bsp	
\label{lastpage}
\end{document}